\documentclass[article]{emulateapj}
\newcommand{\gsim}{\mbox{$\stackrel {>}{_{\sim}}$}} 
\newcommand{\lsim}{\mbox{$\stackrel {<}{_{\sim}}$}} 
\newcommand{\gal}{IRAS 04296}
\slugcomment{accepted 27 Aug 10 AJ}
 
\shorttitle{First Views of IRAS 04296+2923}
\shortauthors{Meier et al.} 
\begin{document} 
 
\title{First Views of a Nearby LIRG: Star Formation and Molecular Gas in 
IRAS~04296+2923}

\author{David S. Meier\altaffilmark{1,2}, Jean
 L. Turner\altaffilmark{3}, Sara C. Beck\altaffilmark{4}, Varoujan
 Gorjian\altaffilmark{5}, Chao-Wei Tsai\altaffilmark{3} \& Schuyler D. 
 Van Dyk\altaffilmark{6}}

\altaffiltext{1}{Department of Physics, New Mexico Institute of Mining and 
Technology, 801 Leroy Place, Socorro, NM 87801; dmeier@nmt.edu}
\altaffiltext{2}{Adjunct Assistant Astronomer, National Radio Astronomy 
Observatory, P O Box O, Socorro, NM, 87801}
\altaffiltext{3}{Department of Physics and Astronomy, UCLA, Los Angeles, 
CA 90095--1562; turner;cwtsai@astro.ucla.edu}
\altaffiltext{4}{Department of Physics and Astronomy, Tel Aviv University,  
69978 Ramat Aviv, Israel; sara@wise.tau.ac.il}
\altaffiltext{5}{Jet Propulsion Laboratory, 4800 Oak Grove Boulevard, 
MS 169-327, Pasadena, CA 91109; varoujan.gorjian@jpl.nasa.gov}
\altaffiltext{6}{Spitzer Science Center, California Institute of Technology, Mail Code 220-6, 
Pasadena, CA 91125; vandyk@ipac.caltech.edu}

\begin{abstract} 

We present a first look at the local luminous infrared galaxy
IRAS~04296+2923.  This barred spiral galaxy, overlooked because of its
location behind the Taurus molecular cloud, is among the half dozen
closest (D= 29 Mpc) LIRGs.  More IR-luminous than either M82 or the
Antennae, it may be the best local example of a nuclear starburst
caused by bar-mediated secular evolution.  We present Palomar J and Pa
$\beta$ images, VLA continuum maps from $\lambda = $20--1.3~cm,
a subarcsecond Keck LWS image at 11.7$\,\mu$m and Owens Valley
Millimeter Array CO(1--0), $^{13}$CO(1--0), and 2.7 mm continuum
images. The J-band image reveals a symmetric barred spiral galaxy.
Two bright, compact mid-infrared and radio sources in the nucleus mark
a starburst that is energetically equivalent to $\sim 10^5$ O7 stars, separated
by $\lesssim 50$~pc. This is probably a pair of young super star clusters, 
with estimated stellar masses of $\sim 10^7~M_{\odot}$ each.  
The nuclear starburst is forming stars at the rate of $\sim12\pm 6~M_{\odot}\,\rm yr^{-1}$, 
or about half of the total star formation rate for the galaxy of 
$\sim 25\pm10~M_{\odot}\,\rm yr^{-1}$. 
IRAS 04296+2923 is very bright in CO, and among the most gas-rich
galaxies in the local universe.  The $^{12}$CO luminosity of the inner half
kpc is equivalent to that of the entire Milky Way. While the most
intense CO emission is extended over a 15\arcsec\ (2 kpc) diameter
region, the nuclear starburst is confined to within
1--2\arcsec\ (150--250 pc) of the dynamical center.  Based on masses
obtained with $^{13}$CO, we find that the CO conversion factor in the
nucleus is higher than the Galactic value, X$_{CO}^{Gal}$ by a factor
of 3--4, typical of gas-rich spiral nuclei.  The nuclear star
formation efficiency is $\rm ^{nuc}M_{gas}/SFR^{nuc} = 2.7\times 10^{-8}~
yr^{-1}$, corresponding to a gas consumption timescale, 
$\tau_{SF}^{nuc} \sim 4 \times 10^7$ yrs. 
The star formation efficiency is ten times lower in the 
disk, with $\tau_{SF}^{disk} \sim 3.3 \times 10^8$ yrs. 
The low absolute star formation efficiency in the disk implies that the 
molecular gas is not completely consumed before it drifts into the nucleus, and is capable 
of fueling a sustained nuclear starburst.  IRAS~04296+2923 appears to be beginning 
a 100~Myr period as a LIRG, during which it will turn much of its 
$6\times10^9~\rm M_\odot$ of molecular gas into a nuclear cluster of stars.
\end{abstract} 
 
\section{Introduction \label{intro}} 
Luminous (L$_{IR}~ \gsim ~ 10^{11}$ L$_{\odot}$) and ultraluminous
(L$_{IR}~ \gsim ~ 10^{12}$ L$_{\odot}$) infrared galaxies are powered
by prodigious amounts of star formation
\citep[][]{JW85,CHYT91,SM96,Get98,LSRMG98,DS98}.  While ULIRGs are
typically associated with mergers \citep{1988ApJ...325...74S}, in
LIRGs, star formation may be generated by interactions or secular
evolution, often related to bars \citep[][]{KK04}.  LIRGs are relatively rare
\citep[][]{SMNDELR87}.  The few LIRGs close enough for their internal
structure to be resolved are valuable targets for the study of the
local physics of the starburst/gas interaction, such as when and where
the star formation takes place within a galaxy.

IRAS 04296+2923 (hereafter \gal) lies behind A$_{V}~ \sim 5.5$ mag of
extinction from the dark cloud L1500 in Taurus (Table \ref{bdata}).
Low-resolution optical and HI spectra allowed it to be identified as a
galaxy \citep[][]{SHDYFT92,CKSYT95}.  \gal\ has an IRAS luminosity of
$9.8 \times 10^{10}~L_{\odot}$ \citep[29 Mpc, for H$_{o}$ = 71 km
  s$^{-1}$ Mpc$^{-1}$][]{SMKSS03}, making it the fifth most
IR-luminous galaxy within 30~Mpc, at the lower bound of the LIRG
class. It is 60\% more IR-luminous than M82, and 35\% more than the
Antennae, exceeded only by NGC~1068, NGC~2146, NGC~7552, and NGC~1365
in the local universe.
   
Little is known about \gal.  We discovered its nuclear starburst in a
subarcsecond mid-infrared imaging survey with the Long Wavelength
Spectrometer (LWS) on the Keck Telescope; 
it has a warm IRAS 60/100$\mu$m color that
is a diagnostic of concentrated star formation.  The bright double
mid-IR source revealed by LWS ($\S$\ref{rirr}), $\lesssim
1$\arcsec\ (150 pc) across, suggested the presence of a new starburst
galaxy in the local universe.

In this paper we present near-infrared (NIR), mid infrared (MIR),
millimeter and centimeter radio continuum images of \gal, as well as
NIR and millimeter spectral imaging.  While its location in Taurus
hinders optical and ultraviolet observations of \gal, infrared and
radio observations are unaffected, and they are the most reliable
probes of star formation in dusty starbursts.

\section{Observations \label{obs}} 

\subsection{Near-Infrared J-Band and Pa $\beta$ Images from the Hale Telescope \label{nir}}

\begin{figure}
\epsscale{1.0}
\plottwo{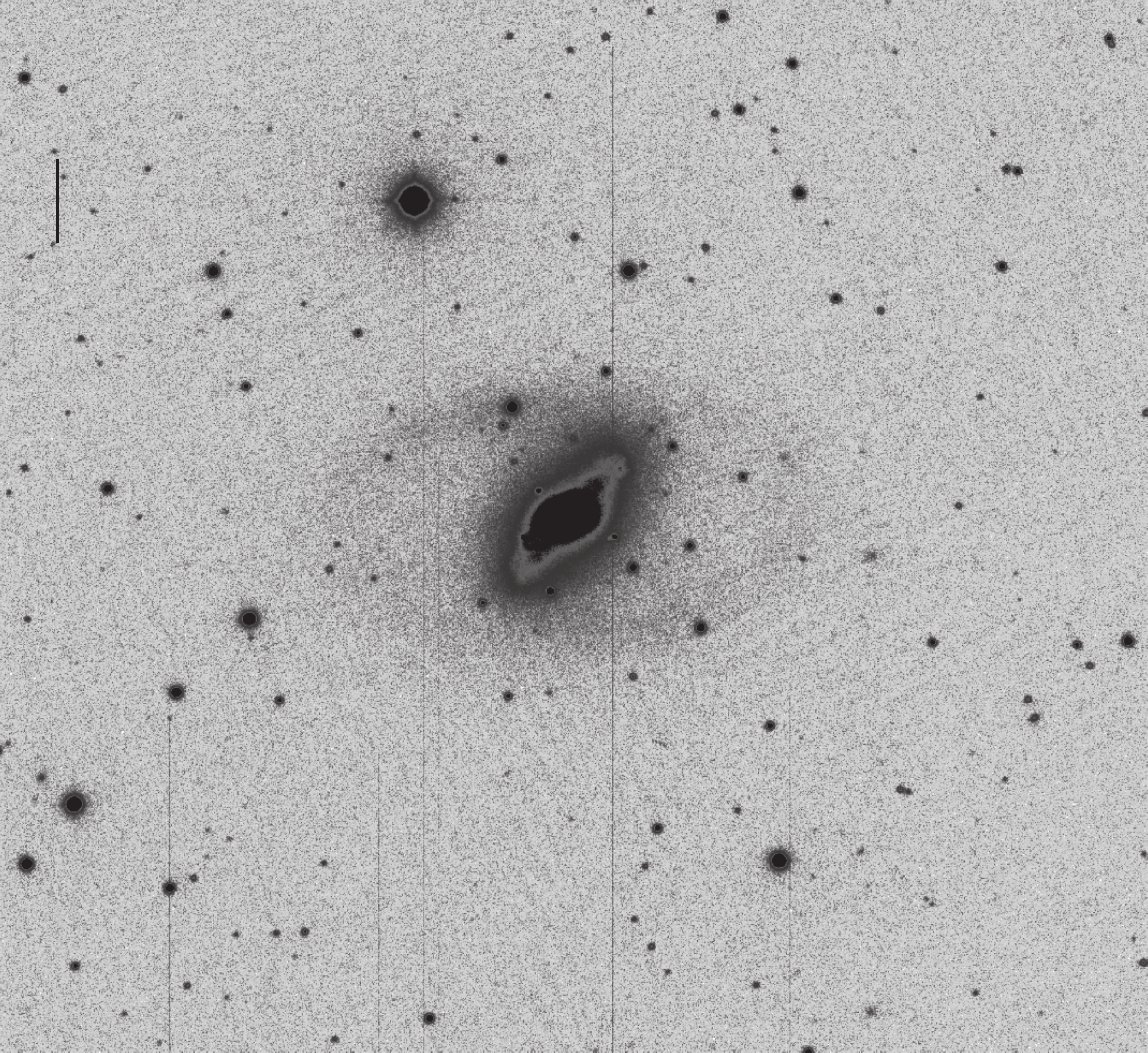}{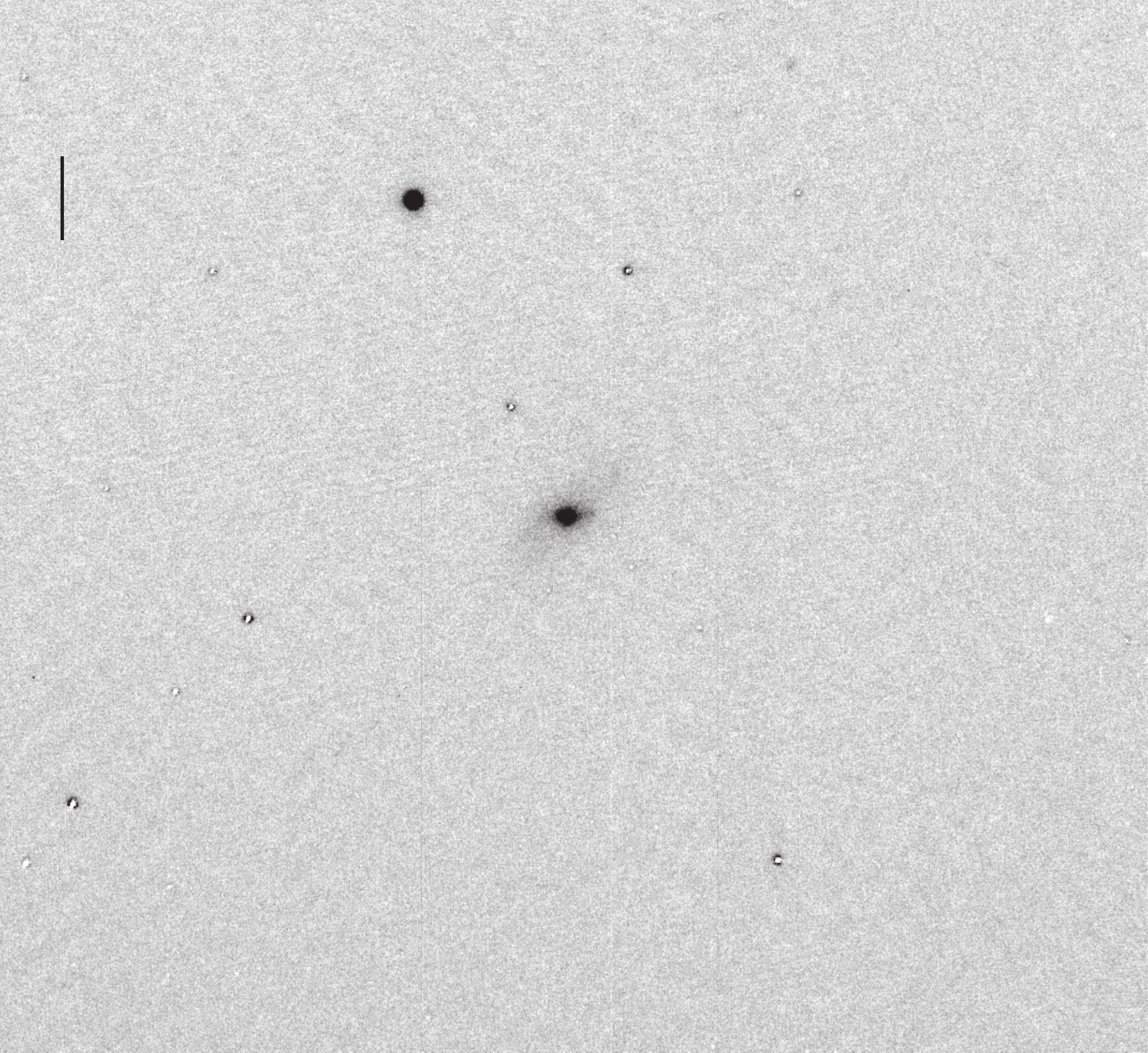}
\caption{{\it Left)} The J band (1.25$\mu$m) image of \gal\ taken on the
Palomar 5-meter. {\it Right)} Palomar continuum-subtracted Pa $\beta$
image of \gal. The many point sources in the image are foreground
Galactic stars of the Taurus region.  Residuals from the continuum 
subtraction are seen in Pa $\beta$; in particular a large residual contribution 
remains from the bright foreground star to the northeast of the 
galaxy.  These residuals are likely due to slight mismatches in the 
point spread function of the (partially) saturated J band stellar images.
The scale bar is 20\arcsec\ (2.8 kpc) in each plane, and the orientation
is north-up, east-left. \label{Jband} }
\end{figure}

\gal\ was observed using the Palomar Hale 5-meter
telescope\footnote{Based on observations obtained at the Hale
  Telescope, Palomar Observatory, as a part a continuing collaboration
  between the California Institute of Technology, NASA/JPL, and
  Cornell University.}  on 2005 January 23 UT with the Wide-field
Infrared Camera \citep[WIRC;][]{WIRC03} at prime focus, in broad-band
J and narrow-band Pa$\beta$ filters.  The camera has a pixel size of
0$\farcs$25.  Exposure times on-source were 11 minutes (3 $\times$ 20
sec frames, 11 pointings) for the continuum band and 18 minutes (3
$\times$ 60 sec frames, 6 pointings) in the emission-line band, with
equivalent off-source exposure times to measure the sky.  The
observing conditions were good with seeing at $1\farcs0$ FWHM.  Median
dark current and filtered sky images created for each band were
subtracted from the frames.  Finally a coadded mosaic was produced
from each set of on-source frames.  We scaled the counts in each
band's mosaic, via aperture photometry of several bright, uncrowded
stars in the field around the galaxy, and subtracted the J-band mosaic
(see Fig.~\ref{Jband}a) from the Pa$\beta$ mosaic, to produce an
approximate continuum subtracted image (see Fig.~\ref{Jband}b).  The
images in the broad-band and narrow band filters were not explicitly
flux-calibrated using observations of standard stars, so we are unable
to derive a value for the net Pa$\beta$ flux in the continuum
subtracted image.

\subsection{Mid-Infrared Images from Keck \label{mir}}

\begin{figure}
\epsscale{0.5}
\plotone{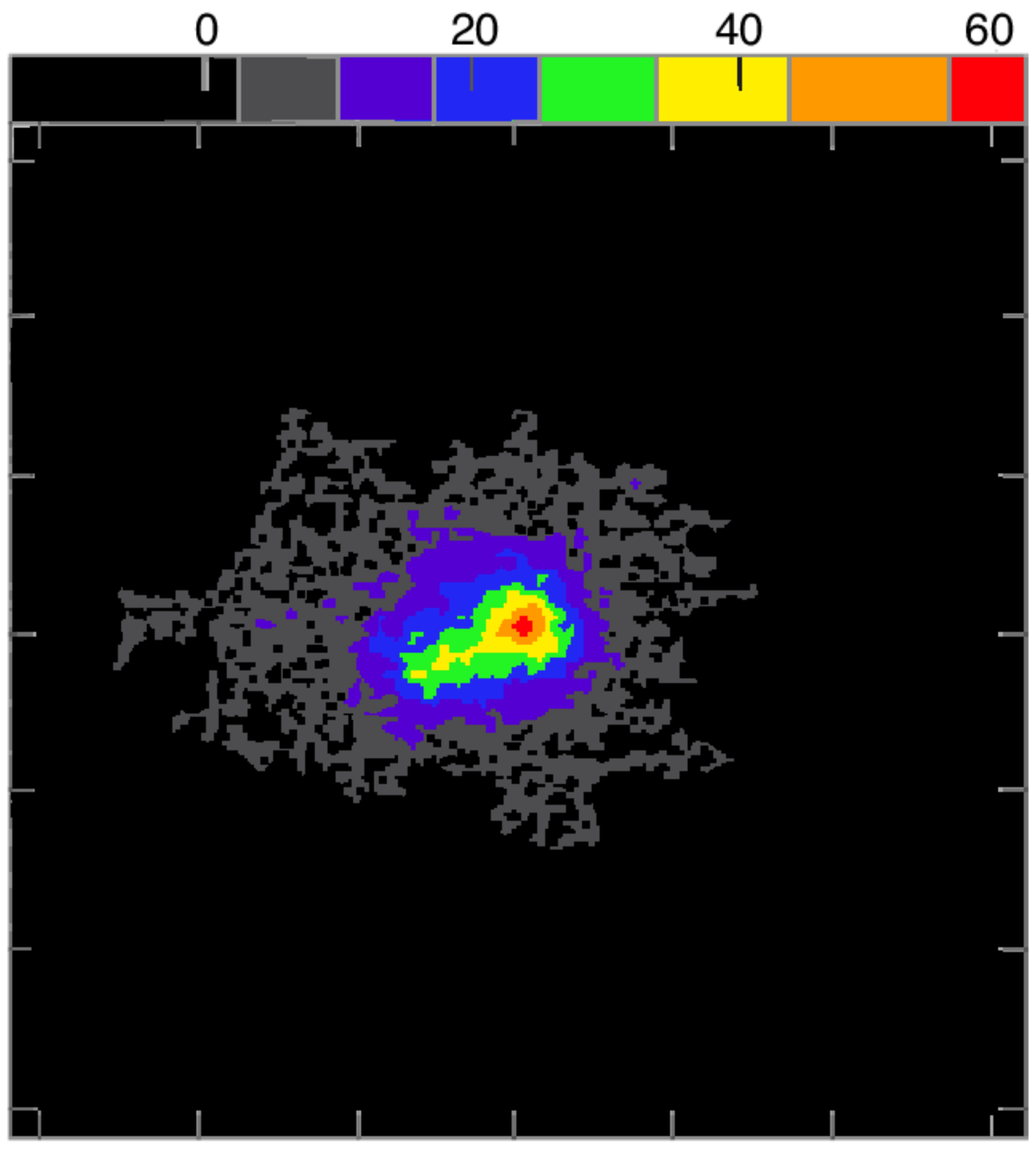}
\caption{The Keck LWS image of \gal\ at $11.7\mu$m. The scale in
counts per pixel is above the figure and the conversion is  
$4.1\times 10^{-5}$~Jy/count for a 360 second integration.
North is up and east is left.
\label{ircont} }
\end{figure}

\gal\ was observed on 24 December 2002 with the Long Wavelength
Spectrometer (LWS) in imaging mode on the Keck I Telescope, with the
standard $11.7\mu$m/SIC filter.  The standard stars
were $\alpha$~Aur, $\alpha$~Aries and $\beta$~Leo, observed both
before and after the source exposures.  High quality images at both
wavelengths were obtained (Fig.~\ref{ircont}); however, the absolute
flux calibration was uncertain due to intermittent instrumental
problems, and so we reobserved the galaxy on 31 January, with the same
parameters as confirmation. Absolute flux calibration is estimated to
be good to 10\%, based on derived flux counts on standard star
before/after observations throughout the night.  Stellar images were
about 5 pixels or $0\farcs4$ FWHM.

\subsection{Radio Continuum Images from the VLA \label{cm}}

The radio data were taken with the NRAO VLA\footnote{The National
  Radio Astronomy Observatory is a facility of the National Science
  Foundation operated under cooperative agreement by Associated
  Universities, Inc.}. There were two projects: AB1092 on 2003 July 7
with the A configuration for high angular resolution at $\lambda=$
20~cm, 6~cm, 3.5~cm, 2~cm, and 1.3~cm (L,C,X U and K bands) and AT309
on 2005 July 5 with the array in BC configuration for moderate
resolution at 1.3~cm only.  3C48 was the primary calibrator for both
runs. The 20, 6, and 3.6~cm data were processed with AIPS, including
baseline corrections. The 2~cm and 1.3~cm data were observed in
fast-switching mode and reduced according to the AIPS prescriptions
for high frequency calibration.  The details of the radio imaging are
discussed below (\S \ref{rirr2})

\subsection{Millimeter CO Lines and Continuum from OVRO \label{mm}}
 
Simultaneous aperture synthesis observations of the $^{12}$CO(1-0)
transition (115.271 GHz) and the $^{13}$CO(1-0) transition (110.201
GHz) of \gal\ were made with the Owens Valley Radio Observatory (OVRO)
Millimeter Interferometer between 2003 November 1 and 2004 May 8.  The
interferometer consisted of six 10.4 m antennas with cryogenically
cooled SIS receivers \citep[][]{OVRO94}.  Observations in ``Compact'',
``Low'' and ``High'' configuration were obtained, with system
temperatures (single sideband) ranging from 220 - 450 K at 115 GHz.
64$\times$4 MHz channels were used to cover both transitions, giving a
velocity resolution of 10.5 km s$^{-1}$ for CO(1-0).  The phase center
is within $0\farcs5$ of the 2 $\mu$m peak listed in Table \ref{bdata}
and the adopted systemic velocity (LSR) was 2100 km s$^{-1}$.
Simultaneous low spectral resolution, wideband (128$\times$31.25 MHz)
COBRA observations were used to generate a 3 mm continuum image
($\nu_{o}~ \simeq$ 111 GHz), free from $^{12}$CO(1-0), CN(1-0) and
$^{13}$CO(1-0) contamination.  The data were calibrated using the MMA
software package.  Phase calibration was done by observing the quasar
J0336+323 every 25 minutes.  Absolute flux calibration was done using
Uranus as primary flux calibrator and 3C273 and 3c454.3 as secondary
flux calibration, and is good to $\sim$10\%.

Mapmaking was done in MIRIAD and subsequent data analysis and
manipulation was done with the NRAO AIPS package.  Maps 
were generated with natural weighting to maximize sensitivity, uniform 
weighting to maximize resolution and robust = 0 weighting to simultaneously 
optimize sensitivity and resolution.  Unless stated the analysis results from the 
naturally weighted data.  Integrated intensity images are straight moment~ 0 
maps with all emission brighter than 1.3 $\sigma$ (1$\sigma$ for $^{13}$CO) 
per channel included.  The OVRO primary
beam is $\sim$64$^{''}$ at 115 GHz.  Corrections for the primary beam
have not been applied so emission at the edge of the primary beam will
be somewhat underestimated.  Structures extended on scales larger than
$\sim$52$^{''}$ for CO(1-0) will be resolved out by the
interferometer.  No single-dish observations of this galaxy exists, so
no estimate of the amount of resolved-out flux is possible for the
mm-wave images.  However, 52$^{''}$ corresponds to spatial scales 
of 7.3 kpc and given that the emission must be extended on this scale 
in an individual channel to be missed, we consider it unlikely that 
significant flux is resolved out.  If large amounts of flux are missed then the 
observed molecular richness is a lower limit.

\section{Results: A First Look at IRAS 04296+2923 \label{res}}

Near-infrared images of \gal\ reveal a normal, barred spiral galaxy in
starlight. The bright, concentrated nuclear starburst is prominent at
MIR and radio continuum wavelengths.  In $\S$\ref{rirr} we discuss the
large field Palomar near-IR images of the galaxy and its Pa$\beta$
emission, and the Keck high resolution $11.7\mu$m MIR image.  In
$\S$\ref{rirr2} we discuss the VLA radio images, which were mapped in
four ways, one to match beams for extended emission dominated by
synchrotron emission, and two for matching compact emission dominated
by free-free from HII regions.  In \S \ref{mols} we discuss the CO and
molecular gas.

\subsection{Infrared Views \label{rirr}} 
The Palomar J band continuum image and a narrow band
continuum-subtracted image of Pa$\beta$ in \gal\ are shown in
Figure~\ref{Jband}.   
The near-infrared continuum at J-band reveals the
stellar population.  \gal\ is an inclined, barred spiral with outer
ring-like spiral arms (SBb/c(r)). The total extent of the emission is
roughly 2-2.5\arcmin, or 17-20~kpc (8-10 kpc in radius).  The stellar
distribution appears bi-symmetric and undistorted.  The $Pa\beta$
image shows line emission to be very concentrated and bright in the
galactic nucleus, with some very weak emission along the inner arms.
The compact nature of the bright Pa$\beta$ emission shows that the
starburst is highly concentrated at the nucleus. Since the images are
not photometric, and extinction is probably variable across the
galaxy, we cannot precisely quantify the relative brightnesses of disk and
nuclear star formation from the Pa$\beta$ image. 

Mid-IR continuum emission traces warm ($\sim$ few hundred K) dust
heated by young stars in the starburst.  The high resolution
(0.3\arcsec) mid-IR continuum image at 11.7$\mu$m made with LWS on
Keck of \gal\ is shown in Figure \ref{ircont}; the field of view of
this image is only 10\arcsec\ $\times$ 10\arcsec, or about 1.4 kpc on
a side.  A similar $18.75\mu$m image (not shown) will be published in a 
subsequent paper (Turner et al. 2010, in prep.).
The 11.7$\mu$m filter contains two PAH features; however,
based on observations of the starburst galaxies NGC~7714 and Arp~220
\citep{2004ApJS..154..188B,2007ApJ...656..148A} we estimate that at
most 15\% of the 12$\mu$m continuum is due to PAH emission. 

The mid-IR emission comprises a bright source, 
which cannot be resolved at the 0\farcs 3
diffraction limit of Keck at 11.7$\mu$m, and a secondary source or
tail extending to the east of the main source, separated from the main
source by $\sim$1\farcs5 ($\sim$200 pc).  The total flux density is   
 $S_{11.7\mu m}=680$~mJy.  These two compact, mid-infrared sources
account for roughly half of the IRAS $12\mu$m flux of the entire
galaxy.

\subsection{Radio Views \label{rirr2}}

\begin{figure}
\begin{center}
\includegraphics*[width=.4\columnwidth,angle=0]{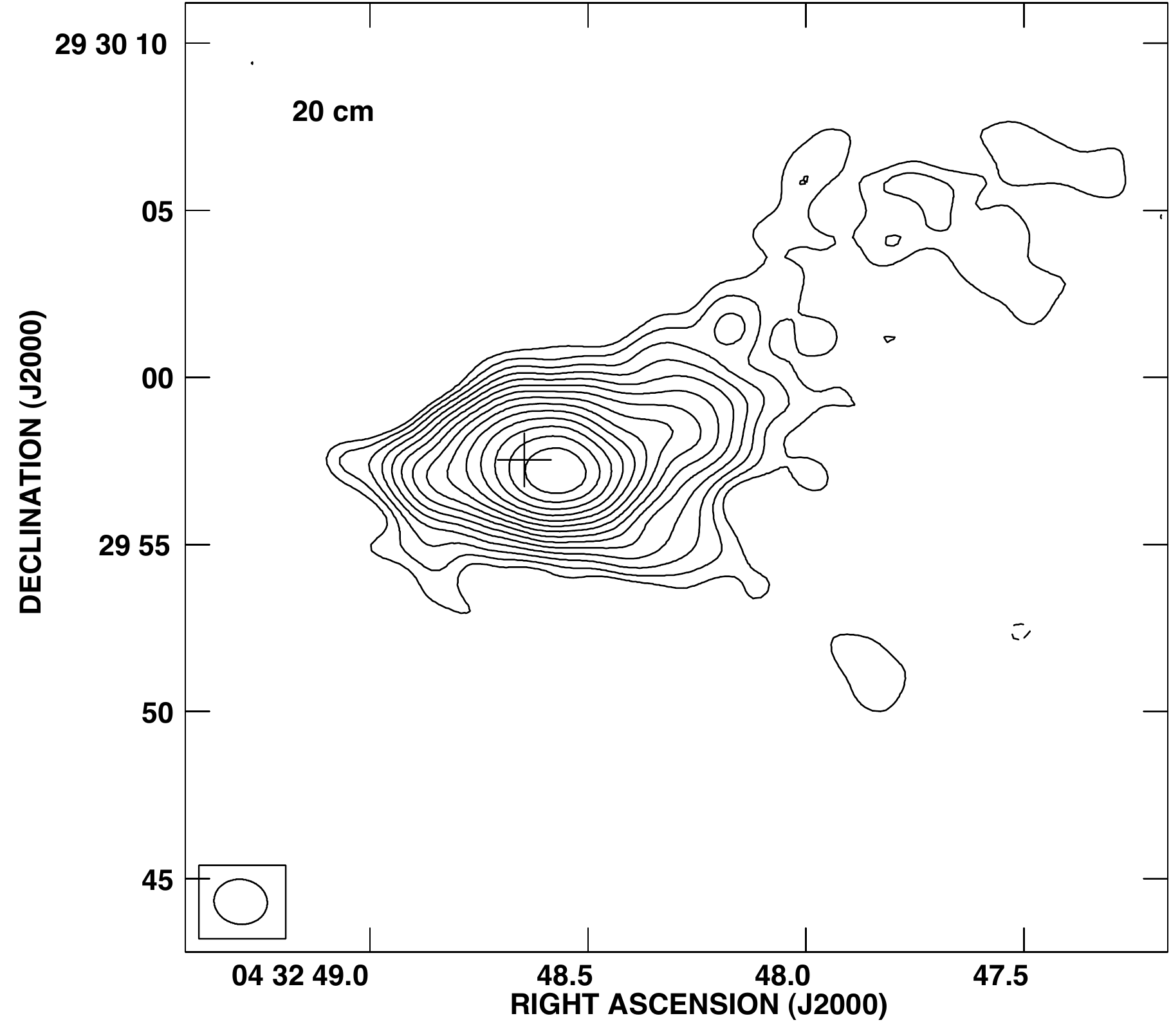}
\includegraphics*[width=.4\columnwidth,angle=0]{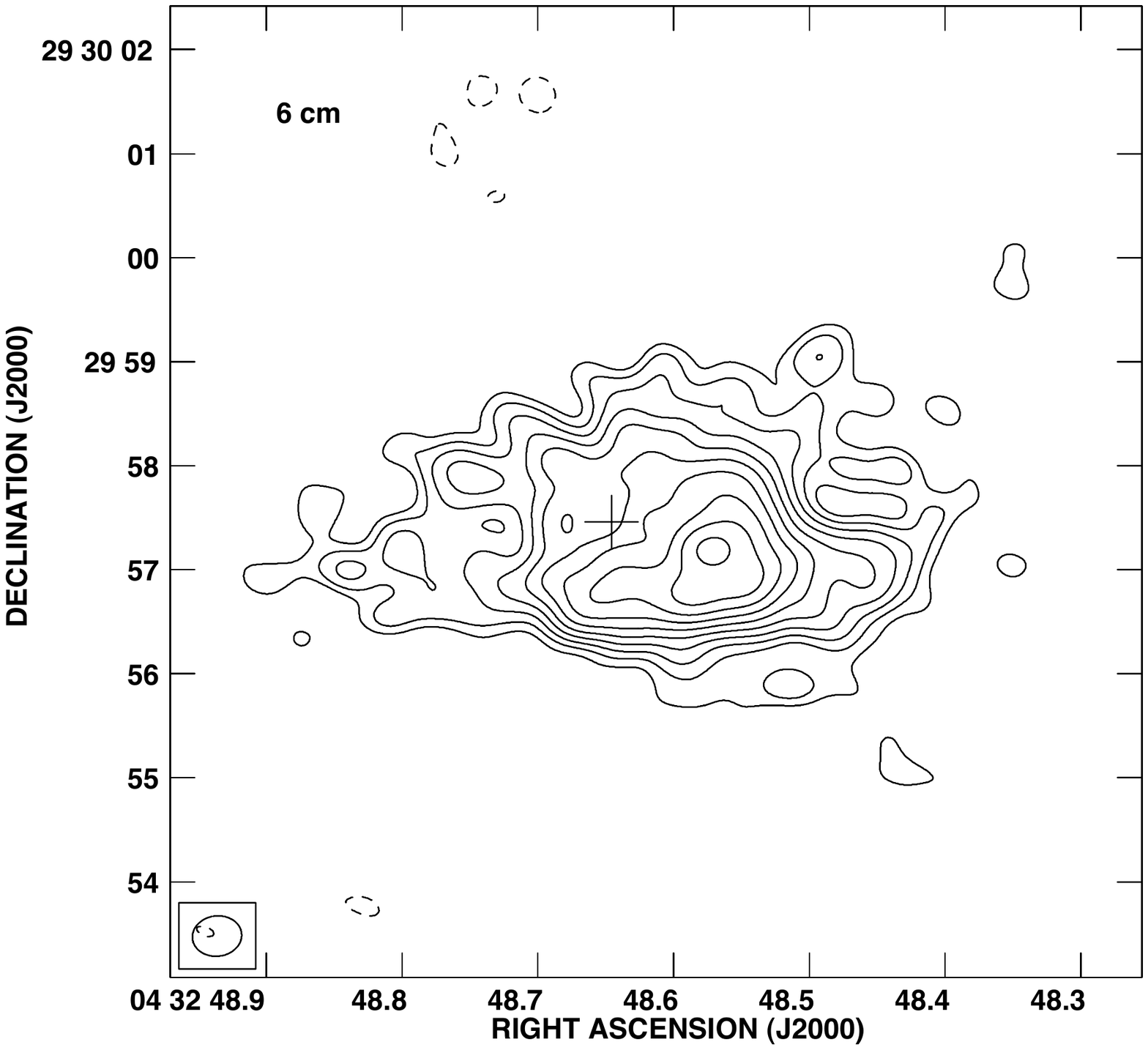}
\end{center}
%\bigskip
\begin{center}
\includegraphics*[width=.4\columnwidth, angle=0]{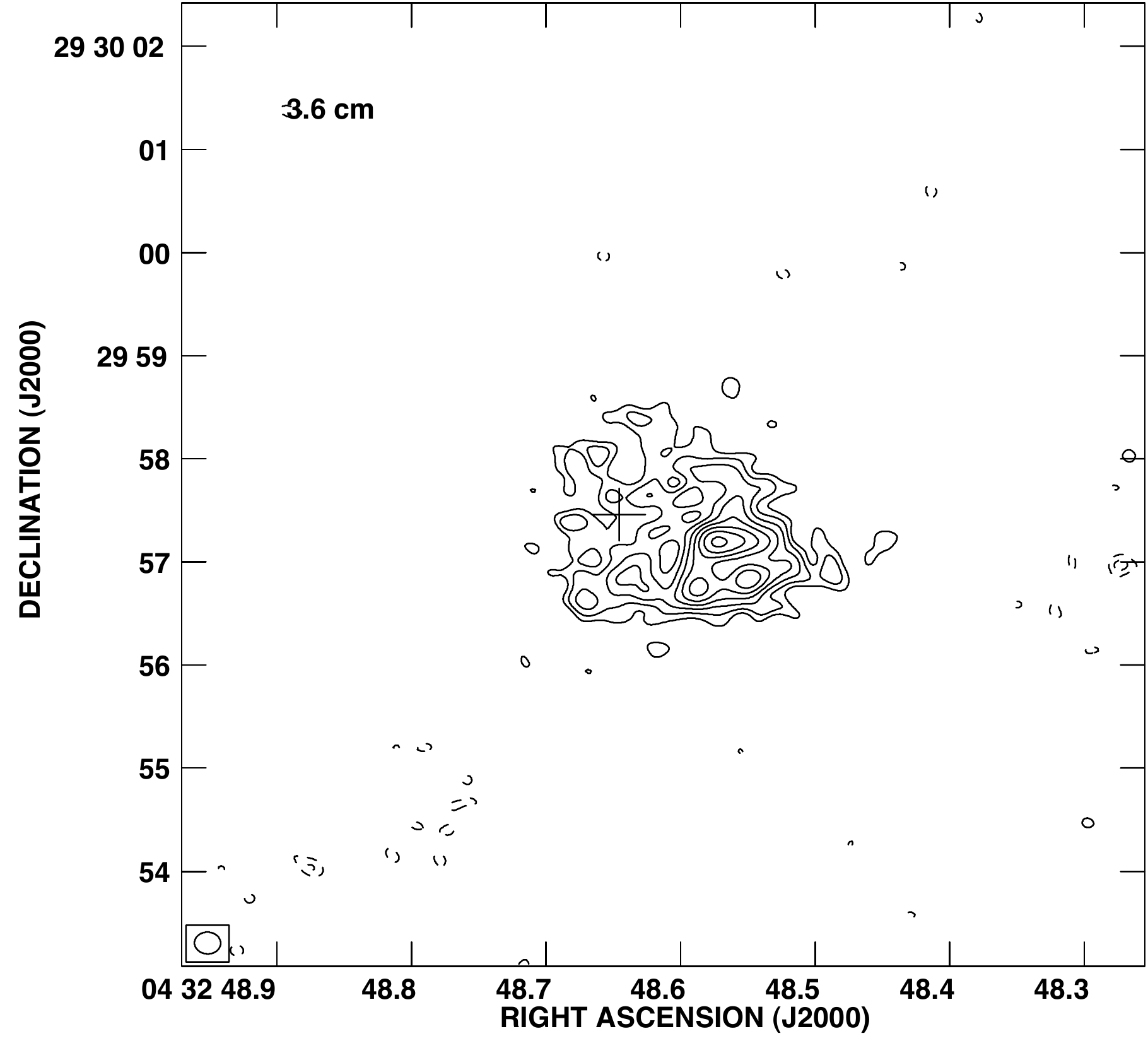}
\includegraphics*[width=.4\columnwidth, angle=0]{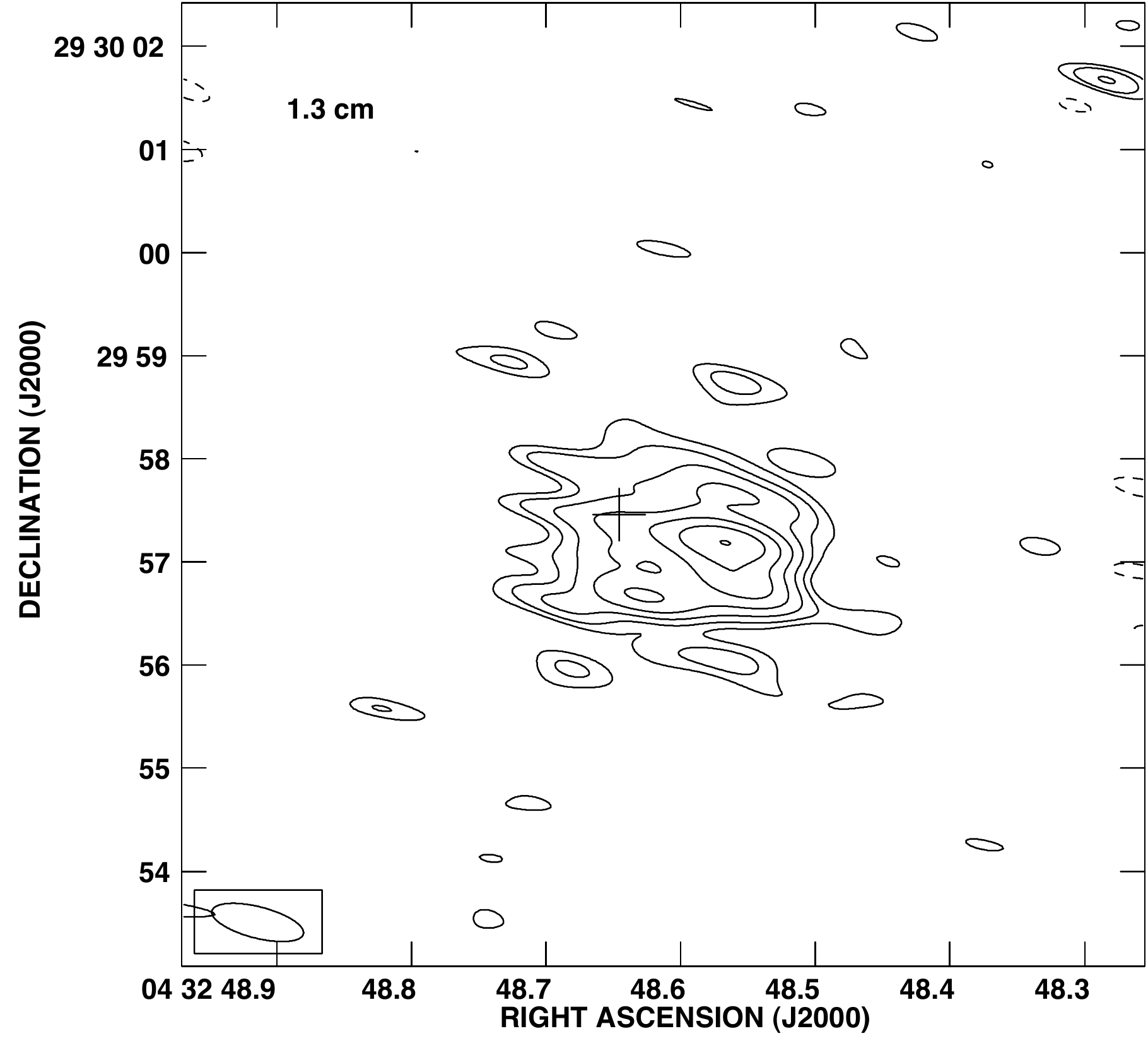}
\end{center}
\caption{Maps of the radio continuum, emphasizing extended
emission.  {\it (top left):} 20 cm. {\it (top right)} 6 cm. {\it (bottom left):} 3.6 cm. {\it (bottom
right):} 1.3 cm with the BC array.  Contour intervals are $\pm2^{n/2}$
($3\sigma$) of the following increments: 0.25 mJy, 0.15 mJy, 0.12 mJy, and 0.25
mJy/bm for 20, 6, 3.6 and 1.3 cm, respectively. The dynamical center
is displayed in the 20 cm and 6 cm maps.
\label{mapsi}}
\end{figure}

VLA maps of \gal\ across the cm waveband are presented in Figure
\ref{mapsi}, with fluxes and noise values given
in Table \ref{Trc1}.  The radio continuum emission is a combination of
thermal bremstrahlung emission from the H{\textsc II} regions and
nonthermal synchrotron emission from supernovae and their remnants
associated with the starburst.  The brightest radio emission, like the
mid-IR emission, is concentrated within the central
1--2\arcsec\ region of $\sim$ 150--250 pc extent; higher resolution
maps shown below give a better comparison with the Keck mid-IR
images. 

Separating free-free emission from the HII regions from the synchrotron 
emission is possible because these sources have different spectral  
and spatial characteristics. However, complications arise when 
comparing aperture synthesis maps at different frequencies, so the spectral 
analysis must be done with care.  The synchrotron component in galaxies is 
spatially extended with a
spectrum that falls steeply with increasing frequency. Free-free
emission from HII regions is compact with spectra that are either flat
or rising with frequency \citep[e.g.,][]{1994ApJ...421..122T}.
Except at 1.3~cm, the robustly-weighted VLA maps of Figure~\ref{mapsi} were done
with a single array configuration, so the maps have responses to
extended structure that vary with frequency; shorter wavelength maps
are less sensitive to extended emission. The
shortest baselines determine the maximum detectable spatial scales,
$\theta_{max}\sim \lambda/B_{min}$, which are listed in Table \ref{Trc1}. So 
the maps of Figure~\ref{mapsi} cannot be compared directly to obtain spectral
index maps.  Since single dish or lower resolution VLA fluxes do not exist for
\gal\ except at 20~cm, we cannot estimate the effects of missing short baselines 
except at 20~cm.  Our A array maps, with total flux 
$S_{20\,cm}= 140\pm 7$~mJy, recover the entire 20~cm flux
recorded by \citet[][]{CHSS96}, and so the 20~cm map is probably a
good representation of the total radio emission, although there may be
faint extended emission that is undersampled. 
From the 20 cm image we can see that the extended radio emission is aligned
along the apparent bar of the galaxy (\S \ref{mols}).  The bright
mid-IR sources and the peak of the radio emission mark a
starburst located at the inner end of the northwestern arm traced by CO 
(\S \ref{mols}).

Free-free and nonthermal synchrotron sources can be separated by
determining the spectral index, $\alpha$, of the emission ($S\propto
\nu^\alpha$) so long as the beams are matched, not only in beamsize
($\theta_{min}$), but also in $\theta_{max}$, defined by
$u_{min}=B_{min}^{EW}/\lambda,~v_{min}=B_{min}^{NS}/\lambda$.  
Toward this end we have 
made a series of ``cut" maps, which are high-pass spatially filtered
images with common $\theta_{min} $ and $ \theta_{max}$, that enable us
to construct images with nearly identical $(u,v)$ coverages and allow
us to compare the radio spectra for the compact sources, at the cost
of losing response to extended emission. Since the star formation
arises largely in compact sources $\lesssim$1\arcsec\ (150 pc), we expect to
detect nearly all of the free-free emission in the ``cut" maps and
resolve out most of the synchrotron emission, which is our goal.

\begin{figure}
\begin{center}
\includegraphics*[width=0.4\columnwidth, angle=-90]{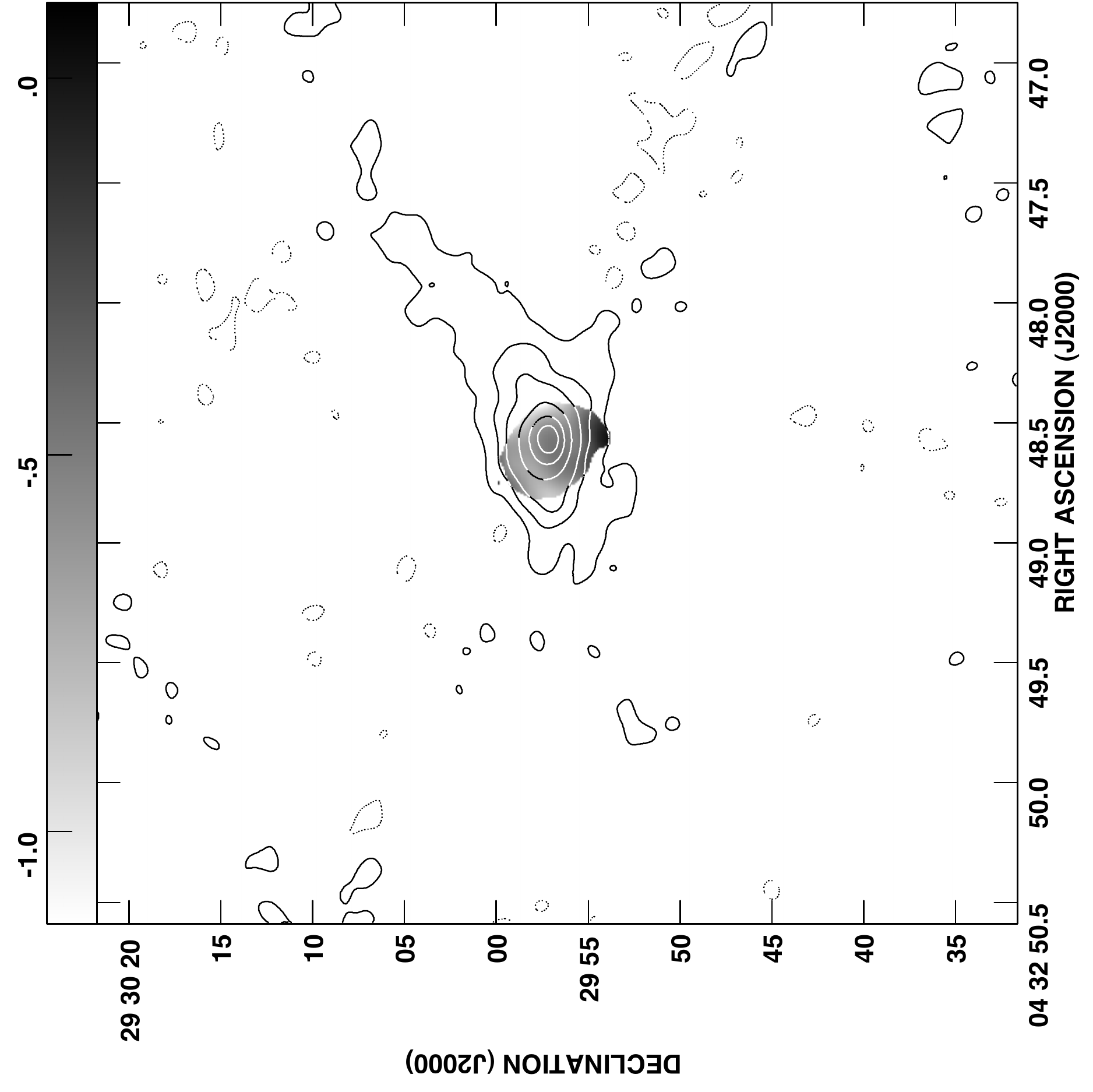}
\end{center}
\caption{The spectral index between 1.3 and 20 cm in greyscale,
overplotted with representative contours of the 20 cm emission.  The
spectral index is not uniform over the whole frequency range covered,
but maps at other frequencies appear similar to this.
\label{spixi} }
\end{figure}

The first set of ``cut" maps were made to match the beams at 20, 6,
3.6, and 1.3 cm, with a consistent $(u,v)_{min}$ of $18k\lambda$
($\theta_{max} \sim 10$\arcsec, 1.4 kpc).  The beam for this
series of maps is 0\farcs82 $\times$ 0\farcs29, p.a. -89\degr. For the
1.3~cm emission, we used the BC configuration data. From the cut maps,
we mapped the spectral index $\alpha$, shown in Figure~\ref{spixi}.
The spectral index is negative over the entire range, confirming that
the extended emission cm-wave emission is largely non-thermal, typical
of large spiral galaxies and LIRGs
\citep[e.g.][]{CHYT91,C92,CVBGSP08}. Fluxes from this set of maps are
measured for a $4.5\arcsec\times3\arcsec$ box (630 pc $\times$ 420 pc)
are shown in Table~\ref{Trc1}.

\begin{figure}
%\epsscale{0.6}
%\includegraphics*[width=\columnwidth]{figure5.pdf}
\begin{center}
\includegraphics[width=5in]{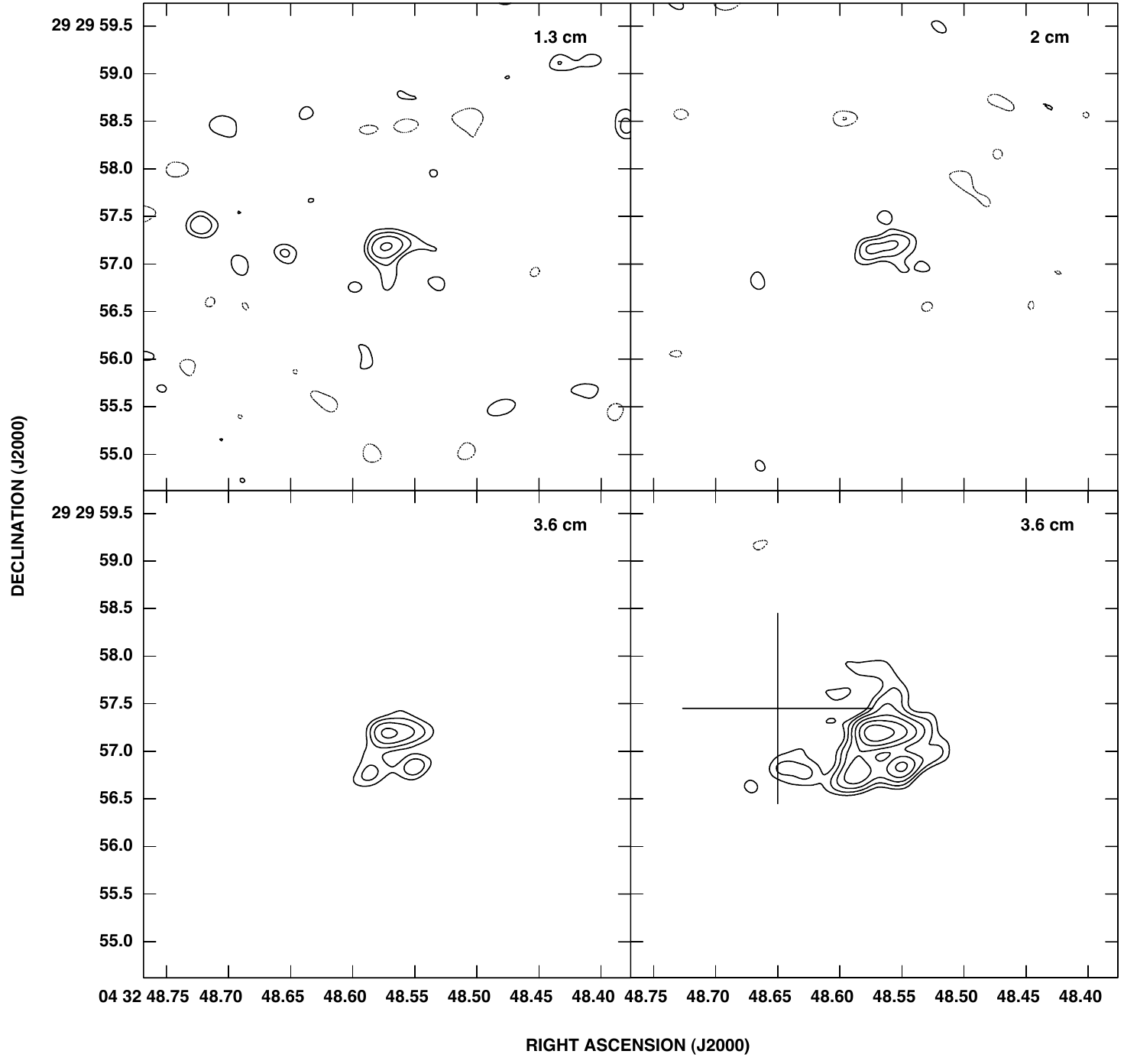}
\end{center}
\caption{Matching 3.6, 2, and 1.3~cm continuum images of \gal\ made
from ``cut" data as described in the text, with consistent $(u,v)$
coverage and a beamsize of 0\farcs15 $\times$ 0\farcs14,
p.a. -66\degr. The bright central source, elongated east-west with a
size of $\sim$0\farcs5 has a flat spectum, containing flux $\sim
3$~mJy. The central arcsecond contains $\sim$6~mJy at 1.3~cm.
Contours are $\pm$2$^{0.5 i} \times 0.5$~mJy/beam
($\sim$2.5--3$\sigma$ at 2 and 1.3~cm).  The 3.6~cm maps are most
affected by the cut of short $(u,v)$ spacings.  {\it a) (top left)}
1.3~cm.  Peak flux 1.5 mJy/beam, total flux 7 mJy.  {\it b) (top
right)} 2~cm.  Peak flux 1.1 mJy/beam, total flux 10 mJy.  {\it c)
(bottom left)} 3.6~cm. Peak flux 1.6 mJy/beam, total flux 10-11 mJy.
{\it d) (bottom right)} 3.6~cm map as in c) except plotted with a
lowest contour of 0.2~mJy/beam. Cross marks the location of the
dynamical center from the CO maps.
 \label{trio} }
\end{figure}

A second set of ``cut" maps were made from the shorter wavelength data
to image the compact emission at 3.6, 2 and 1.3 cm.  For these maps,
$(u,v)_{min}=50\, \rm k\lambda$ ($\theta_{max}\sim$ 4\arcsec, or
$\sim$ 550 pc) and they were convolved to match the 0\farcs
27$\times$0\farcs23, p.a. -90\degr\ beam ($\theta_{min}$) of the
3.6~cm map.  These cut maps are shown in Figure~\ref{trio}, and fluxes
are given in Table~\ref{Trc1}.  The peak flux densities are the same
within observational uncertainties for all three maps, 1.1--1.5
mJy/beam.  The $(u,v)$ restriction has eliminated about two-thirds of
the total 3.6~cm flux as compared to the map of Figure~\ref{mapsi}.
The compact emission that remains in these maps has a nearly flat
spectrum: this is emission from HII regions. These radio images are
similar to the 11.7$\mu$m image of Figure \ref{ircont}, given the
lower resolution (0\farcs3) of the mid-IR image.  We identify the
brightest, elongated radio source with the mid-IR peak, and the
extension visible to the southeast in the 3.6 cm image
(Fig.~\ref{trio}) with the extension in the mid-IR image.

\begin{figure}
%\epsscale{0.9}
%\includegraphics*[width=\columnwidth]{figure6.pdf}
\begin{center}
\includegraphics[width=6in]{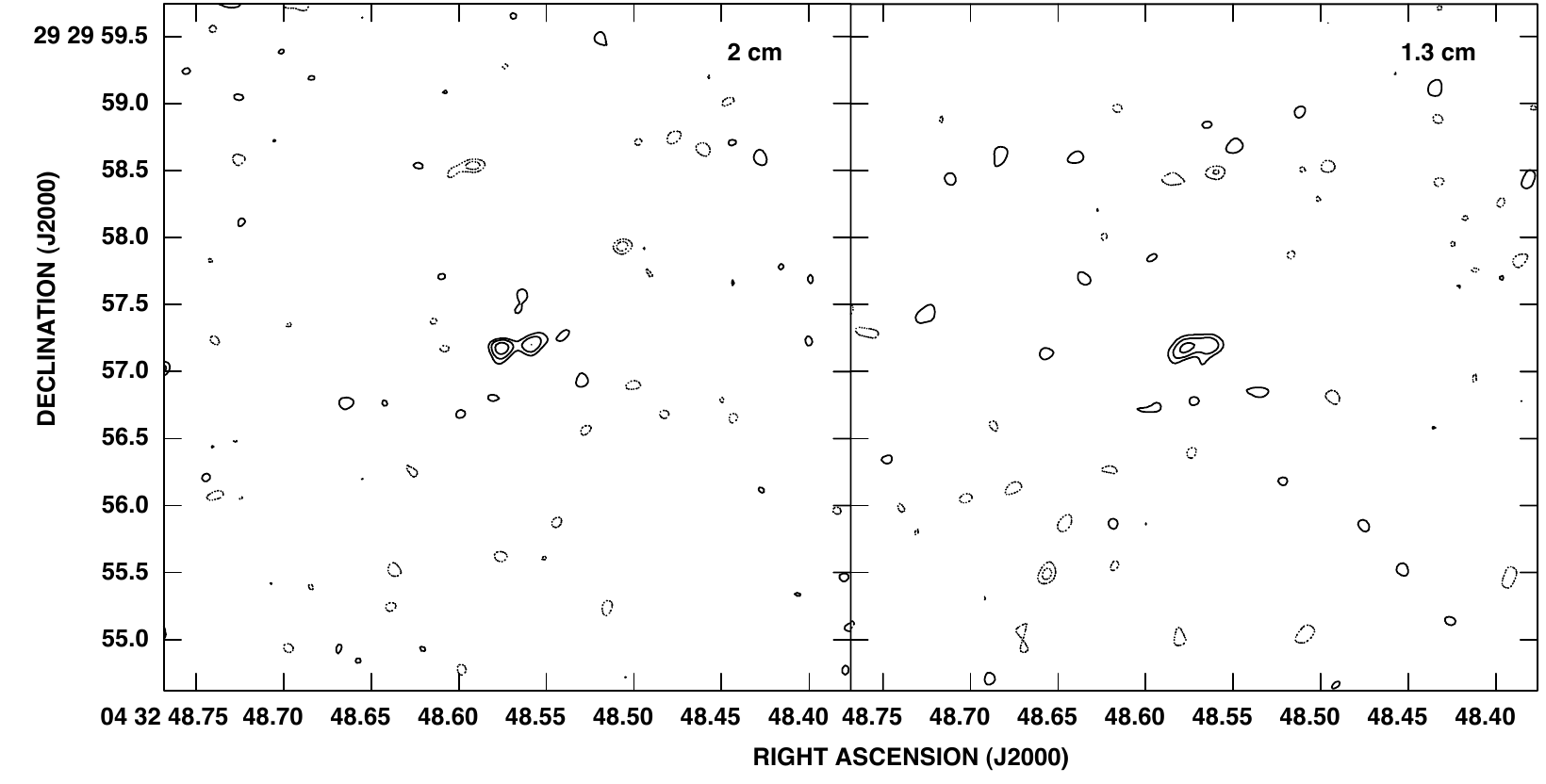}
\end{center}
\caption{ Matching 2 and 1.3~cm continuum images of \gal. Contours are
$\pm$2$^{0.5 i} \times 0.35$~mJy/beam (2.5$\sigma$). Beam for both
maps is 0\farcs15 $\times$ 0\farcs14, p.a. -66\degr.  {\it Left)} 2~cm
image, with $(u,v)$ coverage cut to match the 1.3~cm image as
described in text. Peak flux 0.9 mJy/beam.  {\it Right)} 1.3~cm map
convolved to the same beam as the 2~cm image in a). Peak flux 0.8
mJy/beam.\label{duo} }
\end{figure}

A third set of ``cut" maps, designed to match beams at 2 and 1.3~cm,
give the highest resolution images. Like the previous set of maps,
they are cut to $(u,v)_{min}=50\,\rm k\lambda$ but these maps are
instead convolved to match the smaller 2 cm beam of
0\farcs15$\times$0\farcs14, p.a. -66\degr.  These images resolve the
brightest continuum source (Fig. \ref{duo}) into a double source
separated by 0\farcs2, or 30 pc oriented east-west. The eastern source
is the stronger.  Slight differences in the spatial structure of the
double source at the two frequencies might be due to differences in
azimuthal (u,v) coverage or to 1.3~cm seeing.  This source is flat in
radio spectral index, indicating that the compact emission is largely
free-free emission from HII regions. Comparison of the three sets of 
maps suggests that the total amount of flat-spectrum, compact emission
in the nucleus of \gal\ is $\sim$10~mJy.

Continuum emission at 2.7 mm was detected with OVRO (Figure
\ref{inti12}d).  The peak 2.7 mm continuum intensity is 10$\pm$1 mJy
bm$^{-1}$ ($4.^{''}6\times 3.^{''}7$) with a total detected flux of
11$\pm$2 mJy (Table \ref{Trc1}).  The millimeter continuum emission is
confined to the nuclear starburst, with position and extent consistent
with the (uncut) cm-wave and infrared continuum.  At this wavelength,
potential continuum sources include synchrotron, free-free emission
and dust.  Extrapolation of synchrotron emission all the way from 20
cm to 2.7 mm is dangerous since it is sensitive to the exact spectral
index, but using the 20 cm emission seen in the first `cut' map and
$\alpha=$-0.8 implies a contribution of 2--3 mJy.  Based on our
estimated nuclear gas mass (\S \ref{gas}) we expect that $\sim$1-2 mJy
of the 3 mm continuum is from dust emission.  From this, we conclude
that the millimeter continuum flux associated with star formation is
$\gtrsim$5--7 mJy at 2.7 mm, consistent with the fraction of the flux
from the compact flat spectrum sources seen directly in cm-wave
continuum.

\subsection{Molecular Views\label{mols}}

\begin{figure}
\epsscale{0.9}
\plotone{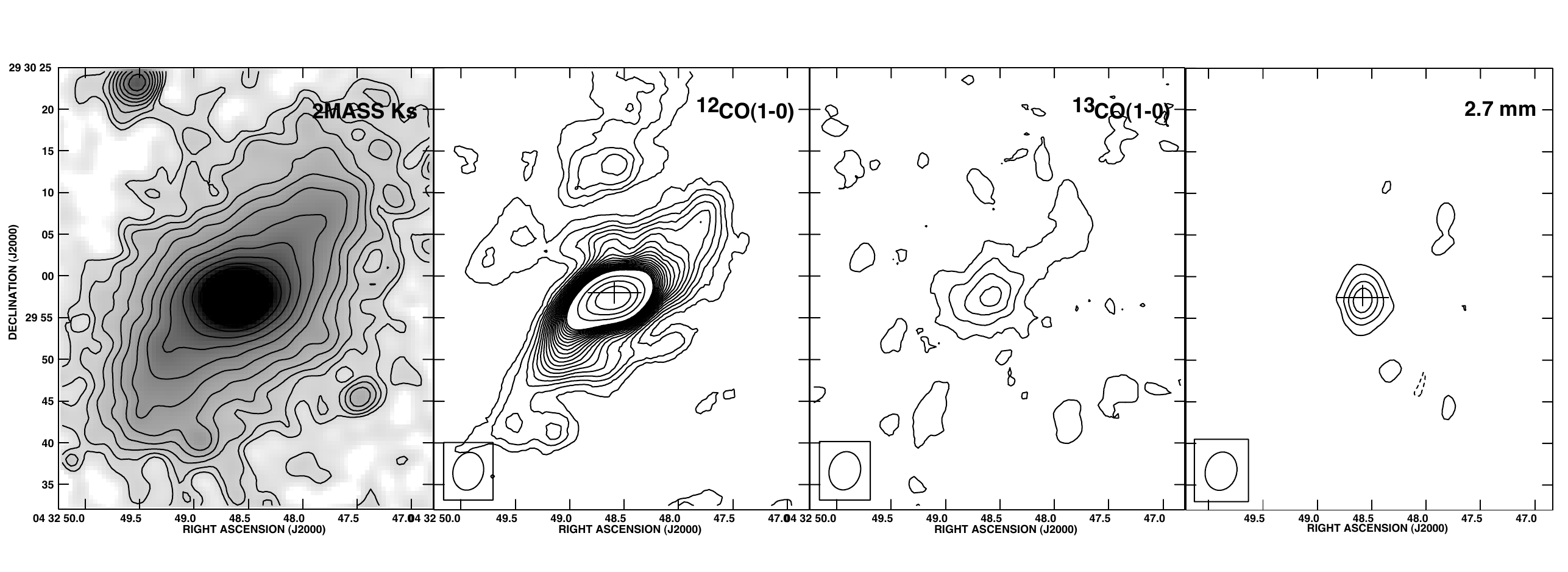}
\caption{{\it Far left)} The 2MASS K$_{s}$ image of \gal\ over the same region
covered by the molecular data.  Contours are logarithmic in arbitrary units. 
{\it Middle Left)} The naturally weighted $^{12}$CO(1-0) intensity map.
Contours are 13.9 K km s$^{-1}$ $\times$1,2,3,...20,30,40,50 for a
beamsize of $4.^{''}6\times 3.^{''}6; -14^{o}$.  {\it Middle right)} The $^{13}$CO(1--0) 
integrated intensity map of \gal\ displayed at the same resolution as $^{12}$CO(1--0).  
Contours are in steps of 10.3 K km s$^{-1}$.  {\it Far right)} The 2.7 mm
continuum map generated from the line free channels of the wide band
dataset.  Contours are in steps of 2.0 mJy beam$^{-1}$ (2$\sigma$) for
a beamsize of $4.^{''}6\times 3.^{''}8;-15^{o}$.  The crosses in the middle 
planes mark the location and extent of the 6 cm radio
continuum emission.
\label{inti12} }
\end{figure}

\begin{figure}
\epsscale{0.7}
\plotone{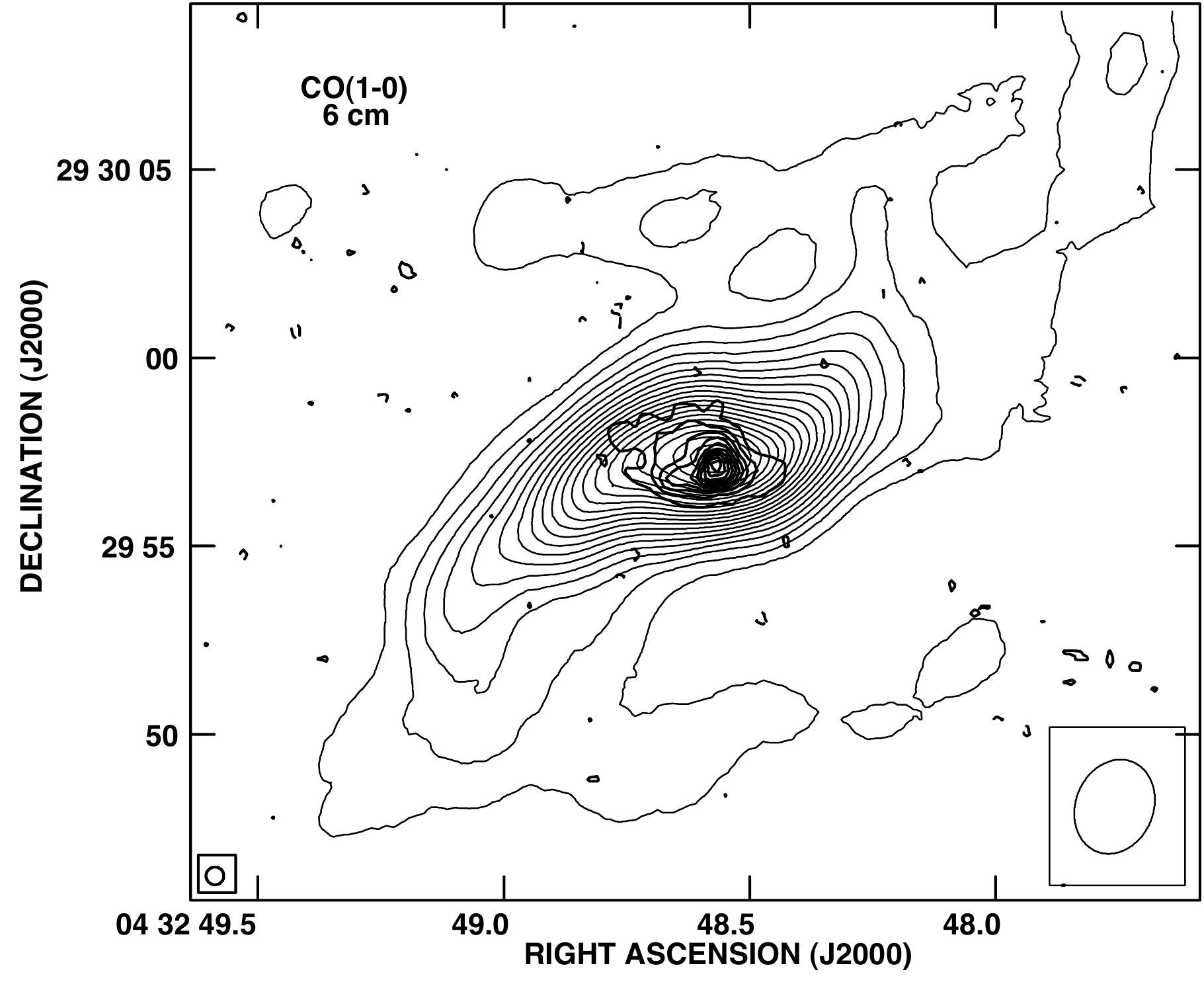}
\caption{6 cm radio continuum (bold lines) overlaid on the central 
region of the robustly weighted CO(1--0) image (thin lines).  CO(1--0) 
contours are in steps of 4 Jy bm$^{-1}$ km s$^{-1}$, while the 6 cm 
contours are 0.2 mJy bm$^{-1}$ $\times$ -3,-1,1,3,5.... The 6 cm 
beam is in the lower left and the CO(1--0) beam ($2.^{''}6\times 2.^{''}1; 21.6^{o}$) 
in the lower right. \label{rc_co}}
\end{figure}

The first observations of molecular gas towards \gal\ are shown in
Figures~\ref{inti12}--\ref{rc_co}.  The CO(1--0) integrated intensity
(Fig.~\ref{inti12}) is remarkably bright, extending over a 2.5 kpc
(diameter) region, peaking at the nucleus.  (When no superscript appears the 
most abundant from, $^{12}$C$^{16}$O is implied; '$^{13}$CO' refers 
to the less abundant $^{13}$C substituted isotopologue.)  
The CO emission extends
well beyond the Pa$\,\beta$ emission region (Figure~\ref{Jband}b) and
the radio continuum (Figure \ref{rc_co}).  Beyond the nucleus the
molecular gas is bar-like, extended southeast-northwest.  Weak
emission is also seen to the north and west of the nucleus.  The CO
morphology closely follows the stellar morphology seen in the Palomar
J band and 2MASS K$_{s}$ images, as shown in Figure \ref{inti12}.

Towards the nucleus, antenna temperatures in the robustly weighted
image (Figure \ref{rc_co}) peak at 10 K averaged over the inner 160 pc
radius. This is a remarkably high brightness temperature for a galaxy at the  
distance of \gal; it implies that the beam filling factor for the CO
must be near unity over regions 300-400 pc in diameter.  The CO line
widths are broadest towards the nucleus but are fairly modest (250 km
s$^{-1}$, FWHM) compared to other luminous starbursts
\citep[][]{SDR92,SDRB97,DS98}, and drop to $\sim$50 km s$^{-1}$
towards the outer molecular arms. The velocity field is consistent
with a rotating disk, except for modest perturbations ($\lsim 50$ km
s$^{-1}$; see below).  We have fit the axisymmetric velocity field
with a Brandt rotation curve; the best-fit parameters are given in
Table \ref{bdata}.  The separate CO component north of the nucleus
follows the same velocity field as does the bar and nucleus.  This
suggests that these clouds are a continuation of the galaxy's disk
(the outer spiral arms).

The molecular morphology is bar-like.  The best fit for the kinematic
major axis from the CO velocity field is at a position angle of
252$^{o}$ (to the receding axis).  This is perpendicular to the long
axis of the CO and IR morphology.  Therefore the observed bar-like
structure seen in CO is not a result of inclination, but is intrinsic.
The strong bar deduced for the stellar distribution makes it
impossible to separately constrain the bar strength and the galaxy's
inclination angle, $i$, over the inner arcminute.  Instead we estimate
the inclination angle from the axial ratio of the outer, presumably
more circular, spiral arms seen in the J band image
(Fig. \ref{Jband}a).  The measurement yields $i = 50 \pm 4^{o}$ and is
consistent with the same p.a. determined from the CO velocity field.
Such an inclination angle implies a peak velocity of $\sim$190 km
s$^{-1}$, or a dynamical mass of $1.6\times 10^{9}~\rm M_{\sun}$ over
the 7$\arcsec$ (R $<$ 500 pc) and $3\times 10^{10}~\rm M_{\sun}$ over the radius, 
$R= 30 \arcsec$ (R $<$ 4.3 kpc). This suggests that \gal\ is similar in size to the Milky
Way at this radius. HI observations at larger radii will be needed to
determine a global dynamical mass for \gal.

To verify that the gas distribution and kinematics are consistent with
response to a bar potential, we generate a simple analytical, weak-bar
model.  These models treat the gas dissipation by adding a damping
term proportional to radial velocity to the standard stellar barred
orbits \citep[e.g.][]{W94,LL94}.  Despite the simplicity of the model,
it matches full hydrodynamical simulations with surprising fidelity
\citep[e.g.][]{LL94}.  The current model is that of
\citet[][]{SOIS99}, except that we use an axisymmetric potential that
generates a Brandt rotation curve.  With the analytical model we can
quickly search barred galaxy parameter space for configurations
consistent with what is observed.

\begin{figure}
\epsscale{1}
\plotone{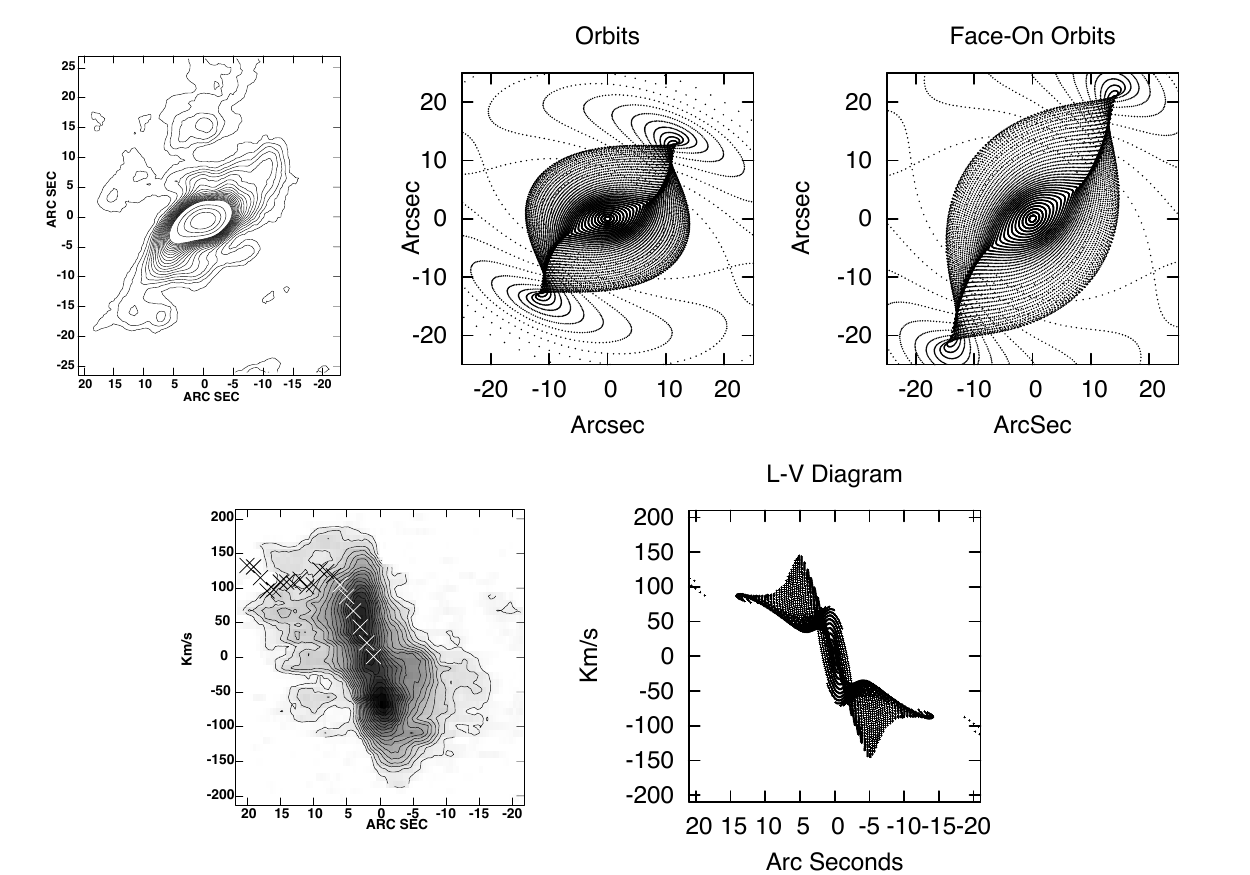}
\caption{An epicyclic weak-bar analytical model for
\gal. {\it Upper left)} The CO(1-0) map (see Figure \ref{inti12}). {\it Upper 
center)} The corresponding model for the morphology of the gas.  Regions of high
dot concentration reflect the expected locations of gas populated
orbits. {\it Upper right)} The same model de-inclined to show the 
face-on morphology of the bar.  {\it Lower left)} The Position-Velocity (PV) 
diagram taken along the major axis (see Table \ref{bdata}) of the galaxy, 
uncorrected for inclination.  The zero velocity coordinate in the figure corresponds 
to 2079 km s$^{-1}$ and the zero position coordinate corresponds to 
$\alpha$(J2000) = 04$^{h}$ 32$^{m}$ 48.$^{s}$56 ;
$\delta$(J2000) = 29$^{o}$ 29$^{'}$ 57.$^{''}$6.  Crosses mark the
fitted azimuthally averaged rotation curve.  Uncertainties are
given by the size of the symbol except for the points at the
largest radii, which have significantly higher uncertainties.  {\it
Lower right)} The PV diagram for the model, generated in a consistent
fashion for comparison with the observed PV diagram.
\label{kine}}
\end{figure}

Figure \ref{kine}, which shows the bar models and rotation curve,
indicates a close agreement of models with the observed morphology and
kinematics. From the best models we find that the molecular gas
distribution of \gal\ is very well reproduced by a large axial ratio
bar.  We assume that the northern portion of the galaxy is the near
side, consistent with two fairly straight trailing arms on the leading
edge of the bar.  The observed velocity residuals have the expected
signatures of inflow along the downstream sides of the bar arms and
weaker outflow on the upstream (so called ``spray" region) sides of
the arms.  Both modeled radial velocities and peak observed velocity
residuals are in the range of V$_{r} \sim 10 - 40 $ km s$^{-1}$.

\section{Discussion \label{disc}}

\subsection{The Star Formation Rate and Luminosity of the Starburst in  \gal 
 \label{starburst}}

From the observed free-free emission ($\S$\ref{rirr2}) we can infer
the Lyman continuum rate of the starburst, its $L_{IR}$, and star
formation rate.  The compact cm-wave emission of \gal\ consists of an
east-west elongated source, centered at RA=04$^h$32$^m$48\fs6,
Dec=29\degr 29\arcmin57\arcsec. The brightest emission corresponds to
a region 0\farcs4$\times$0\farcs2 in size (50~pc $\times$ 30~pc) with
a flux of $\sim$3~mJy and a flat spectrum, located about
1\arcsec\ (140 pc) WSW of the dynamical center.  There is a halo of
emission, stronger to the south of the double source
(Fig.~\ref{mapsi}).  The emission measure, EM, of these HII regions is
high; our high resolution maps (Figs. \ref{trio} and \ref{duo}) detect
only gas with $EM > 10^6$--$10^7~\rm cm^{-6}\, pc$, which is
characteristic of dense or compact HII regions in the Galaxy. The
central double radio source has a total flux $\sim 7$~mJy at 1.3~cm,
and $\sim$10~mJy at 2 and 3.6~cm.  The short wavelength (2 - 0.26 cm)
continuum emission is consistent with star formation having a
free-free flux of $ \sim$5--7 mJy, which implies an ionizing flux from
young stars of $\rm N_{Lyc} \sim 6 \pm 2\times10^{53}~\rm sec^{-1}$
for optically thin emission.  This total Lyman continuum rate is
equivalent to $1.4\pm 1\times10^5$ O7 \citep{MSH05} stars.  Using
STARBURST99 \citep{1999ApJS..123....3L,2005ApJ...621..695V} and a
Kroupa IMF with mass cutoffs of 0.1 and 100 M$_\sun$, we infer a
luminosity associated with the OB stars of $L_{OB}\sim 2 \pm 0.7
\times 10^{10}~\rm L_\sun$ for the nuclear (R$<$150 pc) starburst, to
within a factor of two, given uncertainties in free-free flux and
starburst age.  The mid-IR continuum flux is consistent 
picture: the 12$\mu$m/radio flux ratio of 100 is similar to the values
of 50--100 observed in both Galactic and extragalactic HII regions
\citep[e.g.][]{1982ApJ...255..527G, 1989ApJ...344..135H}.  The thermal
cm-wave fluxes imply a star formation rate of $\sim12\pm 6~M_\sun~
yr^{-1}$ and a mass of $2 \pm 1\times 10^7~\rm M_\sun$ for the nuclear
starburst.  From the morphology, we infer that these are two massive
young super star clusters, ``hyper star clusters", each containing
50,000 O stars and with individual masses of $\sim 10^7~M_\sun$. The
nuclear starburst contains about half of the current global star
formation rate of \gal, based on the ratio of LWS flux for the nuclear
source to the IRAS 12$\mu$ flux for the entire galaxy.

\subsection{A Super CO-Rich Galaxy: Conversion Factors and Estimating \gal\ 's H$_2$ Mass.   \label{gas}}

\gal\ is very bright in CO.  Its total
mapped CO luminosity is L$_{CO}~ = ~1.4 \times 10^{9}~\rm K~km~s^{-1}
pc^{2}$.  The nucleus is the dominant feature in the CO maps.  The
peak integrated CO(1-0) intensity averaged over the central kpc (R =
3.5$^{''}$) is higher than the peak CO brightness averaged over
individual GMC scales in the nearby gas-rich spiral, IC 342
\citep[e.g.][]{MTH00}. Among nearby galaxies, only NGC 2146, also a
LIRG, and possibly NGC 253, can rival it in the richness of its
molecular emission \citep{JH88,YCKRS88}.

The bright CO implies a large H$_2$ mass.  For a Galactic
conversion factor of $X_{CO}^{gal}=2 \times 10^{20}~\rm cm^{-2}(\rm
K~km~s^{-1})^{-1}$ \citep[e.g.][]{Strong,Hunter}, we would derive  a 
molecular gas total mass  of $M_{H_{2}}(R < 500 pc)~=~ 1.6 \times 10^{9}~M_{\odot}$ 
for the central kpc, which is larger than the molecular mass of the entire Milky
Way. This mass is  comparable to the dynamical mass for the
central kpc, as determined by the rotation curve from our model
(Fig. \ref{kine}).   Either molecular gas accounts for the entirety of the dynamical 
mass in the inner 500 pc radius or the CO-to-H$_{2}$ 
conversion factor is different than in the Galaxy. Evidence is accumulating 
that the latter explanation is far more likely in the center of a gas-rich galaxy 
such as \gal. Studies of a variety of galaxies 
indicate that $X_{CO}^{gal}$ as determined from the 
molecular clouds in the disk of our Galaxy is inappropriate for the
central regions of galaxies.  
In (U)LIRGs it is well known that 
the Galactic conversion factor consistently overestimates 
(nuclear) molecular masses by factors of 3-4 
\citep[e.g.][]{DSR93,SDRB97,YSKD03,NGKW05}.  This is often called the
``starburst" conversion factor. However, in fact, this lower conversion factor 
also holds in gas-rich centers of normal spiral galaxies including relatively
quiescent ones such as our own
\citep[e.g.][]{MHWWR96,DHWM98,MT01,WNHK01,Pet01,MT04,MTH08}.  Since
$X_{CO}$ is a dynamical measure of mass, any contribution to the cloud
linewidths from systematic motions driven by the stellar potential
will cause an overestimate of the gas mass. This is particularly
likely in galactic centers, where tidal effects on the clouds become
important \citep[e.g.,][]{MT04}.

Our $^{13}$CO map provides an independent measure of gas mass,
which we can obtain 
from this optically thin line by ``counting molecules".  If we adopt
abundances of [$^{12}$CO/$^{13}$CO] = 60, \citep[e.g.][]{WR94}, and
[$^{12}$CO/H$_{2}$] = $8.5\times 10^{-5}$ \citep[][]{FLW82}, and an
excitation temperature, T$_{ex}$= 20 K, which are typical of clouds in
the nuclei of nearby galaxies, then N(H$_{2}$) column densities can be
estimated from $^{13}$CO line strengths. The peak nuclear column
densities derived from $^{13}$CO are 3.6 times lower than the
corresponding values obtained from $^{12}$CO using the Galactic
$X_{CO}$.  Our $^{13}$CO mass suggests that $X_{CO}^{nuc} \simeq 0.6
\times 10^{20}~\rm cm^{-2}(\rm K~km~s^{-1})^{-1}$ in \gal, similar to
values found for other spiral nuclei, LIRGs, and ULIRGs. 

We conclude that the nuclear H$_{2}$ mass  (including He) for \gal\ is 
$M_{H_{2}}(R< 500 pc)~\simeq ~ 4.3 \times 10^{8}~M_{\odot}$
($\Sigma_{H_{2}}  ~\simeq~ 600 ~ M_{\odot}~\rm pc^{-2}$) for
the inner kpc, with estimated uncertainties of $\sim$50\% due to
unknown $^{13}$CO excitation temperature and abundance.  
Caution should be used in comparing this value 
to masses determined for other galactic centers in
the literature in which the Galactic $X_{CO}$ is used to determine the
mass; those masses are probably systematically overestimated. 

Outside the nucleus, the CO intensity in the disk remains high in 
absolute terms.  Normalized by distance squared, the CO intensity of
\gal\ averaged over its central $45^{''}$ aperture is larger than any
galaxy observed in the FCRAO survey \citep[][]{FCRAO95} out to its 30
Mpc distance.  Over the central arcminute of the OVRO field of view
a total of $M_{gas} (R<30\arcsec) = 6\times 10^{9}~M_{\odot}$ of
molecular gas is implied (for a Galactic conversion factor including
He).  Even adopting the lower M 82 conversion factor \citep[][]{WWS02} for this
rather normal appearing disk, \gal\ has twice the molecular mass of
M~82, including all outflows and streamers.  $^{13}$CO(1--0) is weakly
detected towards the disk in \gal\ so further constraints on the
validity of the conversion factor are testable here as well.  At
10$^{''}$ radii, the values of N(H$_{2}$) derived from $^{13}$CO are
within a factor of 1.6 of what the standard conversion factor
predicts, and along the arms at distances larger than 10$^{''}$
(particularly at the bar ends), the two column densities agree within
the uncertainties.  This suggests that $X_{CO}$ in the disk of
\gal\ may be consistent with Galactic disk values.  However gradients
in excitation temperature, $^{13}$CO opacity and isotopic abundances
can affect the $^{13}$CO mass by at least a factor of two, so this
merits further study.

The IR luminosity of \gal\ is $\sim 9.8 \times 10^{10}~L_{\odot}$
\citep[][]{SMKSS03}.  Averaged over the central arcminute, \gal\ has
L$_{IR}$/L$_{CO}$ = 72 L$_{\odot}~(K~km~s^{-1}~pc^{2})^{-1}$,
L$_{IR}$/M$_{H_{2}}$ = 18 L$_{\odot}$/M$_{\odot}$ and
M(H$_{2}$/M$_{HI}$) $\gsim$ 5, all typical of LIRGs
\citep[e.g.][]{SSSS91,SM96}. Given the rotation curve (\S \ref{res}),
an inclination of 50$^{o}$ and the molecular gas masses derived above,
molecular gas accounts for at least 30\% of the dynamical mass within
the central 500 pc radius (using the lower $^{13}$CO mass) and 15\% of
the dynamical mass over the entire mapped region.

\subsection{Gas Stability Against in the Molecular Bar \label{q}}

Is the molecular gas disk in \gal\ so dense that it will collapse into
new stars without further triggering, or is the disk stable? We can
assess this using the Toomre Q parameter,
\begin{equation}
Q ~ = ~ \frac{\alpha \kappa \sigma}{\pi G \Sigma_{gas}}~ =
\frac{\Sigma_{crit}}{\Sigma_{gas}}
\end{equation} 
where $\kappa$ is the epicyclic frequency \citep[e.g.][]{BT87},
$\sigma$ is the gas velocity dispersion, $\Sigma_{gas}$ is the total
gas surface density and $\alpha$ is a constant of order unity
depending on the structure of the galactic disk (here assumed to be
$\alpha ~=~1$) \citep[][]{S60,T64}. If $Q\lsim1$ the gas disk is unstable.

\begin{figure}
\epsscale{.6}
\plotone{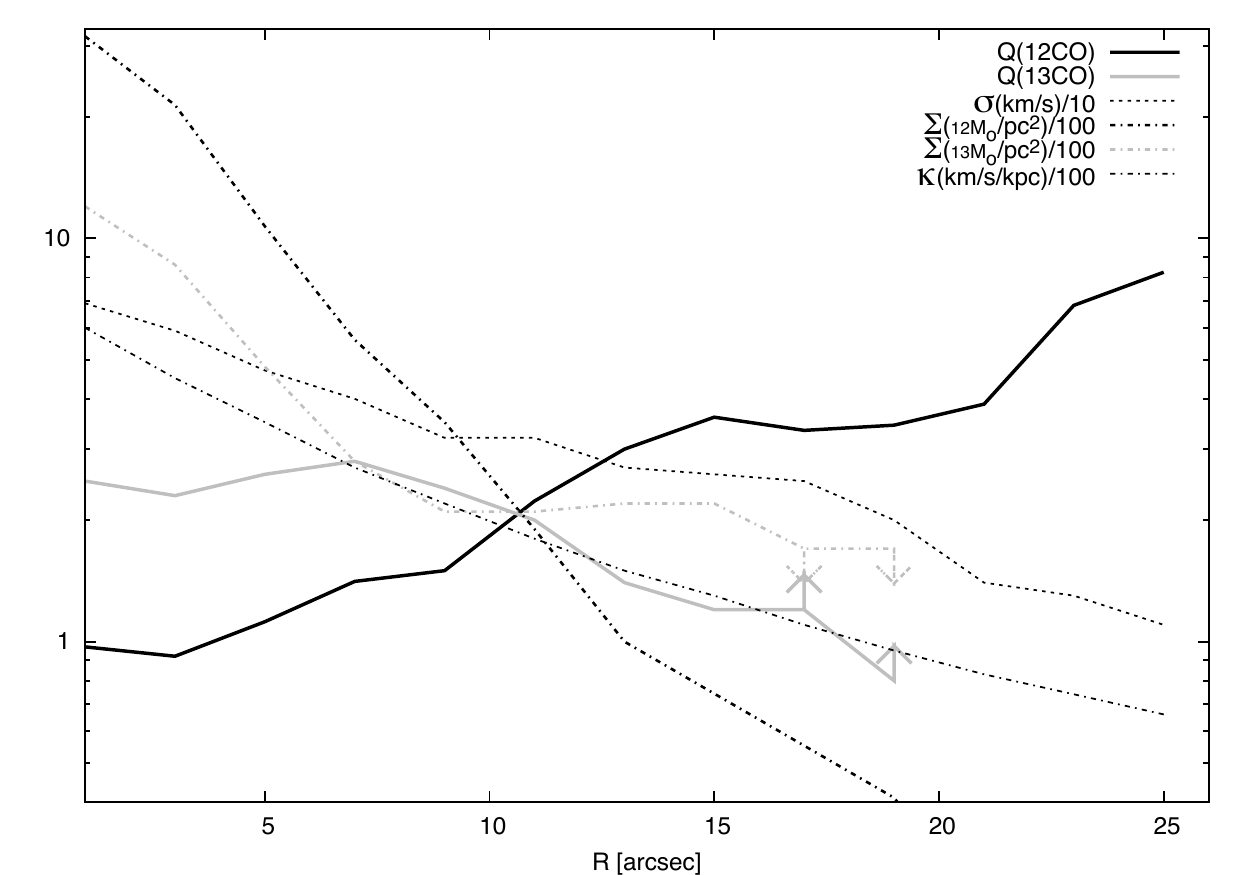}
\caption{The surface densities and Toomre $Q$ parameters for \gal.  The
dot-dashed lines are the azimuthally average molecular surface
densities, $\Sigma_{H_{2}}$, as a function for galactocentric radius.  
Black lines correspond to $\Sigma_{gas}$ calculated using CO(1--0) and 
the Galactic CO-to-H$_{2}$ conversion factor, while gray lines correspond to 
$\Sigma_{gas}$ calculated from $^{13}$CO(1--0) assuming it is optically 
thin (\S \ref{mols}), both inclination corrected.  The velocity disperson $\sigma$ 
(dashed line) is taken from the
line width of the CO(1-0) transition, and $\kappa$ (dotted line) from the fitted
rotation curve (\S \ref{res}).  The resulting Toomre $Q$ parameter 
(\S \ref{q}) is displayed as the thick solid line (black for values from CO determined 
surface densities and gray for $^{13}$CO gas surface density determinations).  Units 
and scaling for the vertical axis are given in the figure legend.
\label{qplot}}
\end{figure}

Figure \ref{qplot} displays the observed values of $\Sigma_{gas}$
(calculated both from $^{12}$CO and $^{13}$CO, neglecting the HI
surface density contribution, which if conservatively estimated to 
be distributed uniformly over the disk contributes much less than 
10\% everywhere within R=17.$^{''}$5), $\sigma$, $\kappa$ and $Q$ as a
function of galactocentric radius.  The data has been averaged in
azimuth assuming a disk inclined at 50$\degr$ to the line of sight.
For $^{12}$CO and the Galactic conversion factor, $Q ~\simeq ~ 1$ for the inner
5$^{''}$ and rises to 3 - 4 by the end of the bar.  This dependence is
driven primarily by the rapid drop in gas surface density just outside
the nucleus together with the fact the velocity dispersion of the gas
remains fairly large ($\sigma ~ \simeq ~20- 40$ km s$^{-1}$) along the
entire molecular bar.  At first glance this is consistent with
gravitational instabilities driving the starburst. But note that this
depends on the overestimated H$_{2}$ mass from the standard conversion
factor.  If we use $\Sigma_{gas}$ derived from optically thin $^{13}$CO
emission, a rather different picture appears. In this case derived $Q$'s would 
be effectively constant with radius over much of  the bar and the
disk would be stable everywhere.  This suggests that the starburst is
not a result of gravitational instabilities in a  disk alone and possibly that the 
large shear/non-circular motions in the strong bar act to inhibit star formation 
in the arms.  The morphology of CO and the starburst's location at the inner 
terminus of the bar (Fig. \ref{rc_co}) argue for bar
induced cloud-cloud collisions as inflowing molecular gas piles up in
the nucleus.

\section{Enhanced Star Formation Efficiency in the Nucleus \label{ensfe}}

From the nuclear mass and disk mass determinations, and
the star formation rates (SFR), we can compare nuclear and disk star
formation efficiencies (SFE).  We consider three representations of 
SFE: 1) the gas consumption timescale, $\tau_{SF}= (\Sigma_{gas}/\Sigma_{SFR})$, 
2) the ratio of the mass of formed stars to total mass available in stars + gas, 
$\eta = \rm [M_{stars}/(M_{gas}+M_{stars})]$, and 3) L$_{IR}$/M$_{gas}$.

We obtain the SFR for disk (0.5 kpc $<$R$<$ 4.3 kpc, not inclination corrected) 
and nucleus (R $<$ 500 pc) in different ways. We estimate the nuclear SFR
from our free-free fluxes for the inner 5\arcsec\ region, as
described in $\S$\ref{starburst} (which are also consistent with 
the 11.7\micron\ LWS flux).  We obtained   
$N_{Lyc}^{nuc}= 6 \pm 1 \times 10^{53}~\rm s^{-1}$, equivalent to
$L_{OB}^{nuc} = 2 \times 10^{10}~ \rm L_\odot$, or a total stellar mass
associated with the young stars of $M_{stars}^{nuc}=2 \times 10^7~M_\odot$,  
and an estimated SFR$^{nuc} =12\rm \,M_\odot\,yr^{-1}$ .  For the disk SFR, 
we cannot use our radio continuum maps, since they are not sensitive to extended  
emission.   To estimate the SFR of the disk, we use the IRAS 12\micron\ flux and 
subtract off the nuclear contribution using our 11.7\micron\ LWS flux. 
About half of the  IRAS 12\micron\ flux originates from the inner 5\arcsec\ of \gal.  
The total disk SFR is only slightly higher than that of the nuclear starburst,
at SFR$^{disk}\sim 13 \pm 5\,\rm M_\odot\, yr^{-1}$, or 2--5 times that of the Milky 
Way. The total predicted luminosity in massive stars (if we assume 
the 12\micron\ flux is excited by OB stars) produces only half of the observed 
total infrared luminosity, which will also have contributions from both older  
populations of stars and regions of exclusively low mass star formation. Our
SFR value for the disk may therefore be an underestimate of the true SFR, but 
only significantly so  if the predominant mode of star formation is exclusively low 
mass, which we view as unlikely. 

To determine the nuclear SFE, we use the nuclear gas mass (from $^{13}$CO 
including He) is $M_{gas}^{nuc}= 4.3 \times 10^8~M_\odot$. The nuclear SFE can then be
represented as $\rm (L_{IR}/M_{gas})^{nuc} \sim (L_{OB}/M_{gas})^{nuc}
\sim 30$--50, with the uncertainty due to the gas mass. This is high,
but within the range found for starburst galaxies
\citep{1991ARA&A..29..581Y}. The surface density of star formation for
this region is then $\Sigma_{SFR}^{nuc}$(R$<$ 500 pc) $= 15\pm9~M_\sun\,
\rm yr^{-1}\, kpc^{-2}$. The surface density of gas, which we assume is
all molecular, is $\rm \Sigma_{gas}(R< 500 pc)=600~M_\sun\,
pc^{-2}$.  This SFE point lies about 
an order of magnitude above the nominal relation for the Kennicutt-Schmidt 
law \citep{K98}, although within the scatter. The gas depletion timescale, 
$\tau_{SF}^{nuc}= 4\times10^{7}$~yrs, 50 times shorter 
than the disk value of $\tau_{SF}=2$~Gyr found for the 23 local spiral galaxies of
the THINGS survey \citep{2008AJ....136.2782L} and more than  20 
times shorter than the $\tau_{SF}\sim$1~Gyr in M51 
\citep{2007ApJ...671..333K,2009ApJ...704..842B} at similar gas surface
densities.  In terms of the percentage of molecular mass converted to stars, 
$\eta^{nuc} \simeq 5\pm 1\%$ over the 500 pc of our CO beam, and is almost
certainly significantly higher on smaller GMC scales. This SFE is
already somewhat high for GMC scales of 50 pc or more, but is
characteristic of SFEs seen in the Galaxy on cluster ($<$10 pc) scales
\citep[e.g.][]{2003ARA&A..41...57L}.  Therefore star formation appears to 
progress at an elevated, but not unusual efficiency in the nucleus.

From the  SFR and molecular gas mass estimated for the disk, we can
compute the disk SFE.  Using the Galactic value of the CO-H$_{2}$ conversion factor 
appropriate for disks, we compute a disk-only molecular
gas mass of $M_{H_2}^{disk}= 4.3 \times 10^9~M_\odot$, which is 10 
times the mass of the nuclear gas. The SFR over the entire disk, by contrast, 
is comparable to that of the nucleus, 
at SFR$^{disk}\sim 13 \pm 5\,\rm M_\odot\, yr^{-1}$,.   
We find $(L_{IR}/M_{gas})^{disk} \sim (L_{OB}/M_{gas})^{disk} \sim 5,$ 
uncertain to about a factor of two largely due to gas mass. This 
is similar, within the uncertainties, to the global Galactic value of 8
\citep[][corrected to our value of $X_{CO}^{Gal}$]{1991ARA&A..29..581Y}. 
The surface density of star formation is then
$\Sigma_{SFR}^{disk}\sim 0.3\pm 0.15~\rm M_\odot\, yr^{-1}\, kpc^{-2}$. The
surface density of gas, which we assume is all molecular in this region, is 
$\Sigma_{gas}^{disk}=75~M_\odot\, pc^{-2}$. 
The disk SFE can also be represented by $\Sigma_{SFR}/\Sigma_{gas}^{disk} =
3.0\times 10^{-9}~yr^{-1}$, or a gas consumption timescale of $\tau_{SF}^{disk}\sim
3.3\times 10^8~$yrs.  The gas consumption timescale is a factor of six shorter
 than the typical THINGS galaxy 
\citep{2008AJ....136.2782L}, and a factor of three shorter than the average
value for M51 \citep{2009ApJ...704..842B}.  In terms of percentage of molecular 
gas converted into stars, $\eta^{disk} = 0.5\%.$  Within the
uncertainties, this is consistent with the SFE of $\sim$1\% observed for
Galactic disk star formation on GMC scales
\citep{2003ARA&A..41...57L,ISK07,HKSK09,Jet09}.  

A low disk SFE is consistent with the model of a nuclear starburst produced 
through secular evolution.  If nuclear star formation is 
seen, it means that fresh gas must be present in the nucleus, else it would  
have formed stars long ago. In a morphologically undisturbed 
galaxy such as \gal, the nuclear gas must be a product of secular evolution. 
If the SFE of the disk were higher, then the gas would be used up before 
it arrived in the nucleus. Much of the molecular gas presently lies within 1.7 kpc 
(12\arcsec) of the center. At this radius the orbital timescale is $\sim 10^8$~yrs.  
If we estimate the velocity for the outermost gas in \gal\ to drift inward via secular 
evolution to be a few km s$^{-1}$ \citep[\S 3.3][]{A92,KK04}, then 
the timescale for all of the CO gas in the disk of \gal\
 to drift into the nucleus is $\sim$1~Gyr. The star formation timescale  
averaged over the disk is $\tau_{SF}^{disk}\sim 3\times 10^8~$yrs. If the SFE were 
much larger than this, then there would be no build-up of nuclear gas, and 
we would not see strong nuclear star formation. 

We conclude that the SFE in the nucleus of \gal\ is an order of magnitude higher than the
SFE in its disk. The value $\eta^{disk}\sim
0.5$\%, for the disk of \gal\ is consistent with the Milky Way value
of $\eta \sim$1\% on GMC scales. The value of $\eta^{nuc} \gtrsim$ 5\% found
for the nucleus is consistent with SFE values on cluster ($<$10 pc)
scales in the Galaxy.  The corresponding star formation timescales
are $\tau_{SF}^{nuc}\sim 3\times 10^7$~yrs for the nucleus and
$\tau_{SF}^{disk}\sim 3.3\times 10^8$~yrs for the disk.
While the disk percolates along at normal
Galactic disk SFRs, the starburst mode ``waits" until the gas has
drifted to the nucleus.

\section{Comparison of \gal\ with other Nearby Galaxies and LIRGs\label{ssb}}

\gal\ is a LIRG of the local universe.  We
now place \gal\ in the context of the other better-known members of that
class: NGC~2146, NGC~7552, NGC~4418, NGC 1365 and NGC 1068.  NGC 1068
and NGC~1365 contain active nuclei which contribute significant
fractions of their total luminosity. NGC~7552 hosts a weak LINER.
NGC 4418 is thought to be a Seyfert 2 because it contains a very a
compact bright nucleus \citep{1990ApJ...365..478E,2003AJ....125.2341E}, but it is 
no more compact than the infrared continuum from \gal . NGC~4418 actually
resembles a compact nuclear starburst more than the composite galaxies
with which it is usually classed.  NGC~2146, along with the less
luminous galaxies M82 and the Antennae, are thought, like \gal, to be
pure starbursts.

In contrast to NGC~4418 and \gal, star formation is more spatially extended
in the remainder of the LIRGs. NGC~7552 and NGC 1068 have kpc-scale
starburst rings, well outside the nucleus. The starburst-dominated galaxies, 
NGC~2146, NGC~1365, M82, and the Antennae, have starbursts extended 
over many hundreds of parsecs.  NGC~7552 and NGC~1365 are the
only other starburst-dominated local LIRG that does not appear overly
disturbed; both have barred stellar appearances similar to \gal's.  NGC
7552 may have experienced a similar dynamical history to \gal, namely
slow, bar-mediated evolution, although in \gal's case the starburst is
concentrated in the nucleus, whereas NGC 7552's enhanced star
formation occurs well before the gas reaches the nucleus.

The local universe LIRGs are a surprisingly diverse class. Many of the
differences are probably because they are at different evolutionary
stages but others appear to be more fundamental. They may perhaps
relate to the underlying causes and mechanisms of the starbursts.

\section{Conclusions: First Look at \gal, A Bar-Driven Starburst Turning On \label{conc}}

The little-studied nearby (29 Mpc) LIRG, \gal, is a barred and
normal-appearing spiral galaxy, with an intense nuclear starburst and
a vast reservoir of molecular gas.   Mid-IR and radio continuum 
imaging show that \gal\ currently has a bright nuclear starburst 
containing the equivalent of $\sim 10^5$ O7 stars ($N_{Lyc}^{nuc} \sim
6\pm 2\times 10^{53}~s^{-1}$). This $\sim$150 pc region alone has the
equivalent of more than half the star-forming luminosity in the entire
500 pc starburst region in M82.  The nuclear starburst region accounts
for $\sim$50\% of the total current star formation in the galaxy as
measured by 11.7\micron\ continuum flux, and contributes 
$\sim$20\% of the total  IRAS luminosity of \gal.  The starburst luminosity is 
concentrated in a double source that is $\lesssim$50~pc in size, located near the
inner end of the northwestern bar arm. Radio and mid-IR images 
suggest that these are the HII regions associated with two massive
young super star clusters containing a total of $2 \pm 1 \times
10^7~M_\odot$ of stars, and a total young stellar luminosity of
$L_{OB}=2\pm 0.7\times10^{10}~L_\odot$.

The morphology and kinematics of the CO emission of \gal\ show a starburst 
that is being fueled by slow gas inflow along a bar.  We can estimate
the history and lifetime of the starburst from the gas content and
inflow velocity.  If the velocity of gas inflow to the nuclear
starburst is the maximum value of the average observed bar-induced
peculiar velocity, $\sim$20 km s$^{-1}$, a net radial inflow of gas
along the arm width ($\sim$750 pc at 10$^{''}$) of 9 M$_{\odot}$
yr$^{-1}$ results (for a Galactic conversion factor relevant to the arms).  For more
expected net inflow rates of a few percent of the orbital velocity
\citep[][]{A92}, $\dot{M}_{inf}~\lsim$ 1 M$_{\odot}$ yr$^{-1}$.  The
star formation rate in \gal\ from the thermal radio continuum is
consistent with a current, nuclear star formation rate of SFR$^{nuc}\sim 12\pm
6$ M$_{\odot}$ yr$^{-1}$ and as much as SFR$^{tot}\sim25\pm 10$ M$_{\odot}$
yr$^{-1}$ total for the galaxy as a whole.  The nuclear star formation is not in equilibrium 
with the net inflow rate and hence the starburst is currently consuming its reservoir of 
nuclear fuel faster than it is being replenished.  

Stellar mass buildup also dictates that the current rate of nuclear star formation
is not sustainable over the long term, that it is a true starburst.
Subtracting the gas mass from the dynamical mass leaves a small nuclear stellar 
mass of $\lsim 1.2 \times 10^{9}$ M$_{\odot}$.  Nuclear star formation also could not
have been occurring at the current $\sim$12 M$_{\odot}$ yr$^{-1}$ rate
for more than the past $\sim 10^{8}$ yrs or the stellar mass would
exceed the observed dynamical mass.  

We also find that the star formation efficiency is a function of position in the
galaxy. We derive a starburst timescale
of $\tau_{SF}^{nuc}=3.7\times10^7$~yrs for the nucleus (R $<$ 500 pc), or 
$\eta^{nuc}$=5\%; the star formation timescale is ten times longer in the disk,
and the SFE is only $\eta^{nuc}$=$\sim$0.5\%.  
The relatively low SFE observed in the disk, where the gas consumption timescale 
is $\sim$5 orbital times, is necessary so that gas drifting into the nucleus along the bar 
orbits would not be consumed before it reached the center.
Any higher value for the disk SFE would be inconsistent with a nuclear starburst 
fueled by gas drifting inward via bar inflow. 

Taken together these facts imply that we are witnessing an early stage of starburst / 
bar inflow induced secular evolution in the history of \gal.  Otherwise the bulge would 
be much larger and the disk depleted of its gas.

\acknowledgements 
 
DSM acknowledges support from the National Radio Astronomy Observatory
which is operated by Associated Universities, Inc., under cooperative
agreement with the National Science Foundation.  The anonymous referee
is thanked for a helpful report. The Owens Valley Millimeter
Interferometer is operated by Caltech with support from the NSF under
Grant AST-9981546.  This research has made use of the NASA/IPAC
Extragalactic Database (NED) which is operated by the Jet Propulsion
Laboratory, California Institute of Technology, under contract with
the National Aeronautics and Space Administration.

{\it Facilities:} \facility{CMA}, \facility{Hale (Wide-field Infrared Camera)}, 
\facility{Keck:I (Long Wavelength Spectrometer)}, \facility{VLA}

\clearpage 
 
\begin{deluxetable}{lcc}
\tablenum{1}
\tablewidth{0pt}
\tablecaption{IRAS 04296+2923 Basic Data}
\tablehead{\colhead{Characteristic}     & \colhead{Value}       &
\colhead{Reference}}
\startdata
Dynamical Center\tablenotemark{a}    & $\alpha(2000) = 04^{h} 32^{m} 48^{s}.65\pm 1^{''}~~~$ &1   \\
(kinematic)  &$ \delta(2000) = +29^{o} 29' 57.^{''}45\pm 1^{''}$   &   \\
2$\mu$m peak (2MASS)     & $\alpha(2000) = 04^{h} 32^{m} 48^{s}.60\pm 0.^{''}3$  &1   \\
  &$ \delta(2000) = +29^{o} 29' 57.''49\pm 0.^{''}3$ & \\
V$_{lsr}$\tablenotemark{a}   &2086 kms$^{-1}$   &1   \\
Adopted Distance  & 29 Mpc  &1   \\
Position Angle\tablenotemark{a} & 252$^{o}$ & 1\\
Inclination & 50$^{o}$ & 1 \\
R$_{max}$(J band) & 8.1 kpc& 1 \\
$R_{max}$(rotation curve)\tablenotemark{a} & 2.15 kpc & 1\\
$v_{max}$\tablenotemark{a} & 190 km s$^{-1}$ & 1\\
$n$\tablenotemark{a} & 1.25 & 1 \\
$\theta_{bar}$\tablenotemark{b} & -85$^{o}$ & 1 \\
$\Omega_{bar}$\tablenotemark{b} & 43 km s$^{-1}$ kpc$^{-1}$ & 1 \\
M$_{H_{2}}$($<$ 500 pc)\tablenotemark{d}  & $4.3 \times 10^{8}~M_{\odot}$& 1 \\
$\Sigma_{H_{2}}$($<$ 500 pc)\tablenotemark{d} & 550 M$_{\odot}$ pc$^{-2}$  &1 \\
M$_{dyn}$ ($<$ 500 pc)\tablenotemark{a} & $1.6\times 10^{9}~M_{\odot}$ &  1 \\
M$_{H_{2}}$($<$4.3 kpc)\tablenotemark{c}  & $5.9 \times 10^{9}~M_{\odot}$ & 1 \\
M$_{dyn}$ ($<30$ kpc)\tablenotemark{a} & $3.0\times 10^{10}~M_{\odot}$ &  1 \\
$\rm M_{HI}^{tot}$ & $1.38 \times 10^{9}~M_{\odot}$& 3 \\
IRAS 12$\mu$, 25$\mu$, 60$\mu$, 100$\mu$& 1.39, 5.90, 42.1, 48.3 Jy &2\\
L$_{IR}$  & $9.8 \times 10^{10}~L_{\odot} $&  2 \\
\enddata
\tablenotetext{a}{Based on the best fitting Brandt rotation curve (\S \ref{res}).}
\tablenotetext{b}{Based on the best fitting bar model (Figure \ref{kine}).}
\tablenotetext{c}{Assuming the standard CO conversion factor (\S \ref{gas}).}
\tablenotetext{d}{From $^{13}$CO(1--0) (\S \ref{gas}).}
\tablerefs{(1) This paper; (2) Sanders et al. (2003); (3) Chamaraux 
et al. (1995)} 
\label{bdata}
\end{deluxetable}

%\clearpage 
\begin{deluxetable}{lccccccc}
\tabletypesize{\scriptsize}
\tablenum{2}
\tablewidth{0pt}
\tablecaption{Infrared and Radio Fluxes}
\tablehead{
\colhead{$\lambda$}
&\colhead{r.m.s.}
&\colhead{Beam (Robust=0)}
&\colhead{$\theta_{max}$}
&\colhead{Peak Flux}
&\colhead{Total Mapped\tablenotemark{a,b}}
&\colhead{Cut Flux\tablenotemark{b,c}}
&\colhead{Cut Flux\tablenotemark{b,c}}
\\
\colhead{}
&\colhead{}
&\colhead{}
&\colhead{}
&\colhead{}
&\colhead{}
&\colhead{$(u,v)_{min}=18\rm k\lambda$}
&\colhead{$(u,v)_{min}=50\rm k\lambda$}
\\
\colhead{}
&\colhead{mJy/bm}
&\colhead{\arcsec$\times$\arcsec, pa$^\circ$}
&\colhead{\arcsec}
&\colhead{mJy/bm}
&\colhead{mJy}
&\colhead{mJy}
&\colhead{mJy}
\\
\colhead{}
&\colhead{}
&\colhead{}
&\colhead{}
&\colhead{}
&\colhead{}
&\colhead{}
&\colhead{}
}
\startdata
20~cm                  & 0.11    & 1.60$\times$1.35, 82.8&38    & 32   & 130 $\pm$ 10 & 71 $\pm$ 5 & \dots \cr
6~cm                    &0.04     & 0.48$\times$0.39, -83.3   & 10   & 3.8  & 59 $\pm$ 5  & 28 $\pm$ 3  & \dots \cr
3.6~cm                 & 0.034 & 0.25$\times$0.21, 88.2  & 7     & 1.5  & 25 $\pm$ 5  & 14 $\pm$ 2  &11 $\pm$2 \cr
2~cm                    & 0.085 & 0.18$\times$0.138, -53.9 & 4     & 1.0  & 10 $\pm$ 1  & \dots& 10 $\pm$ 2 \cr
1.3~cm                 & 0.068 & 0.1$\times$0.09, -80.39   & 2     & 0.5  & 7 $\pm$ 1.5  &  \dots & 7 $\pm$ 1.5 \cr
1.3~cm BC array &  0.16 & 0.91$\times$0.30, 76 & ~20 & 3.7  & 23 $\pm$ 5   & 18 $\pm$ 3 &\dots\cr
2.7~mm               & 0.75   & 4.6$\times$3.8, 8                & 52   & 9.9  & 11 $\pm$ 2 &\dots &\dots \cr
11.7~$\mu$m &5.1 cts/s& 0.4 &n.a.& 71$\pm$10 & 680 $\pm$ 100& \dots & \dots \cr
\enddata
\label{Trc1}
\\
\tablenotetext{a}{ Based on ``uncut" maps with full native A configuration $(u,v)$ coverage. }
\tablenotetext{b}{Quoted uncertainties in the fluxes are based on 3 times the rms times 
$\sqrt{N}$ times the number of beams corresponding to the source size; they do not include resolved-out flux,
or the estimated 5\% (10\% at 2 and 1.3 cm) absolute flux scale uncertainty. Large beam 
fluxes do not exist for wavelengths other than 20~cm, so we cannot estimate
amounts of resolved-out flux except at 20~cm, where it is negligible.  Fluxes are measured within 
$\lambda/B_{min}$ for each map.}
\tablenotetext{c}{Based on ``cut" maps restricted in $(u,v)$ coverage.  
The 18k$\lambda$ fluxes correspond to the maps of Fig.~\ref{mapsi}, and the 
50k$\lambda$ fluxes to those of Figs.~\ref{trio} and \ref{duo}. See \S \ref{rirr2}.} 
\end{deluxetable}

\clearpage 

%\clearpage 

\clearpage

%\clearpage 

%\clearpage

%\clearpage

%\clearpage 

\clearpage
 
\end{document}